\documentclass{aa}  
\usepackage[applemac]{inputenc} 
\usepackage{graphicx}
\usepackage{txfonts}
\begin{document}
   \title{QUBIC: The QU Bolometric Interferometer for Cosmology}



\author{The QUBIC collaboration \and
	E.~Battistelli \inst{5} \and
	A.~ Baù \inst{6} \and
	D.~ Bennett \inst{12} 
	L.~ Bergé \inst{3} \and
	J.-Ph.~ Bernard \inst{2} \and
	P.~ de Bernardis \inst{5} \and
	A.~ Bounab \inst{2} \and
	É.~ Bréelle \inst{1} \and
	E.F.~ Bunn \inst{10} \and
	M.~ Calvo \inst{5} \and
	R.~ Charlassier \inst{1} \and
	S.~ Collin \inst{3} \and
	A.~ Cruciani \inst{5} \and
	G.~ Curran \inst{12} \and
	L.~ Dumoulin \inst{3} \and
	A.~ Gault \inst{9} \and
	M.~ Gervasi \inst{6} \and
	A.~ Ghribi \inst{1} \and
	M.~ Giard \inst{2} \and
	C.~ Giordano \inst{5} \and
	Y.~ Giraud-Héraud \inst{1} \and
	M.~ Gradziel \inst{12} \and
	L.~ Guglielmi \inst{1} \and
	J.-Ch.~Hamilton \inst{1} \and
	V.~Haynes \inst{7} \and
	J.~Kaplan \inst{1} \and
	A.~Korotkov \inst{8} \and
	J.~Landé \inst{2} \and
	B.~Maffei \inst{7} \and
	M.~Maiello \inst{13} \and
	S.~Malu \inst{11} \and	
	S.~Marnieros \inst{3} \and
	S.~Masi \inst{5} \and
	A.~Murphy \inst{12} \and
	F.~Nati \inst{5} \and
	C.~O'Sullivan \inst{12} \and
	F.~Pajot \inst{4} \and
	A.~Passerini \inst{6} \and
	S.~Peterzen \inst{5} \and
	F.~Piacentini \inst{5} \and
	M.~Piat \inst{1} \and
	L.~Piccirillo \inst{7} \and
	G.~Pisano \inst{7} \and
	G.~Polenta \inst{5,14,15} \and
	D.~Prêle \inst{1} \and
	D.~Romano \inst{5} \and
	C.~Rosset \inst{1} \and
	M.~Salatino \inst{5} \and
	A.~Schillaci \inst{5} \and
	G.~Sironi \inst{6} \and
	R.~Sordini \inst{5} \and
	S.~Spinelli \inst{6} \and
	A.~Tartari \inst{6} \and
	P.~Timbie \inst{9} \and
	G.~Tucker \inst{8} \and
	L Vibert \inst{4} \and
	F.~Voisin \inst{1} \and
	R.A.~Watson \inst{7} \and
	M.~Zannoni \inst{6} 	
}

   \offprints{J.-Ch. Hamilton, APC, Paris {\tt hamilton@apc.univ-paris7.fr}}

   \institute{
   	APC, Universit\'{e} Paris Diderot-Paris 7, CNRS/IN2P3, CEA, Observatoire de Paris, 10, rue A. Domon \& L. Duquet, Paris, France.
	\and
	Centre d'\'{E}tude Spatiale des Rayonnements, CNRS/Universit\'{e} de Toulouse, 9 Avenue du colonel Roche, BP 44346, 31028 Toulouse Cedex 04, France.
	\and
	Centre de Spectroscopie Nucl\'{e}aire et de Spectroscopie de Masse, UMR8609 IN2P3-CNRS, Univ. Paris Sud, b\^{a}t 108, 91405 Orsay Campus, France.
	\and
	Institut d'Astrophysique Spatiale, Universite Paris-Sud, Orsay, 91405, France
         	\and
          Dipartimento di Fisica, Universit\`{a} di Roma “La Sapienza”, Roma, Italy. 
	\and 
	Dip. di Fisica ”G.Occhialini” - Universit\`{a} degli Studi di Milano-Bicocca Piazza della Scienza, 3 - 20126 Milano, Italy.
	\and
	School of Physics and Astronomy, The University of Manchester, Alan Turing Building, Oxford Road, Manchester M13 9PL, UK.
	\and
	Brown University, Providence, RI 02912, USA.
	\and 
	University of Wisconsin-Madison, Madison WI 53706, USA.
          \and
	Physics Department, University of Richmond; Richmond, VA 23173, USA.
	\and
	Raman Research Institute, Sadashivanagar, Bangalore 560 080, India.
	\and
	Department of Experimental Physics, National University of Ireland Maynooth, Maynooth, Co. Kildare, Ireland.
	\and
	Università degli Studi di Siena - Rettorato, Via Banchi di Sotto 55, 53100 Siena ITALY
	\and
	ASI Science Data Center, c/o ESRIN, via G. Galilei, I-00044, Frascati, Italy
	\and
	INAF,Osservatorio Astronomico di Roma, via di Frascati 33, I-00040, Monte Porzio Catone, Italy
             }

   \date{Received ; accepted }

 
  \abstract
   {One of the major challenges of modern cosmology is the detection of B-mode polarization anisotropies in the Cosmic Microwave Background. These originate from tensor fluctuations of the metric produced during the inflationary phase. Their detection would therefore constitute a major step towards understanding the primordial Universe. The expected level of these anisotropies is however so small that it requires a new generation of instruments with high sensitivity and extremely good control of systematic effects.}
   {We propose the QUBIC instrument based on the novel concept of bolometric interferometry, bringing together the sensitivity advantages of bolometric detectors with the systematics effects advantages of interferometry.}
   {The instrument will directly observe the sky through an array of entry horns whose signals will be combined together using an optical combiner. The whole set-up is located inside a cryostat. Polarization modulation will be achieved using a rotating half-wave plate and the images of the interference fringes will be formed on two focal planes (separated by a polarizing grid) tiled with bolometers.}
   {We show that QUBIC can be considered as a synthetic imager, exactly similar to a usual imager but with a synthesized beam formed by the array of entry horns. Scanning the sky provides an additional modulation of the signal and  improve the sky coverage shape. The usual techniques of map-making and power spectrum estimation can then be applied. We show that the sensitivity of such an instrument is comparable with that of an imager with the same number of horns. We anticipate a low level of beam-related systematics thanks to the fact that the synthesized beam is determined by the location of the primary horns. Other systematics should be under good control thanks to an autocalibration technique, specific to our concept, that will permit the accurate determination of most of the systematics parameters.}
   {}


   \maketitle
%

\section{Introduction}
This article describes the proposed QUBIC experiment, a Bolometric Interferometer designed to put tight constraints on the Cosmic Microwave Background B-mode polarization anisotropies. These odd parity polarization anisotropies are generated by primordial gravitational waves (and by lensing of even parity polarization at small scales). Detection of these waves would represent a major step towards understanding the inflationary epoch that is believed to have occurred in the early Universe. Tensor modes (primordial gravitational waves) in the metric perturbation are indeed a specific prediction of inflation. The measurement of the corresponding B-mode polarization anisotropies would therefore be a smoking gun for inflation. A detection would reveal the inflationary energy scale, which is directly related to the amplitude of this signal.
The tensor to scalar ratio r is however expected to be small (smaller than 0.2 from today’s best indirect measurement - the contribution of the tensor modes to the temperature and E-mode polarization anisotropy) so that the quest for the B-modes is a major experimental challenge. Such a small signal (at least an order of magnitude below the E-mode signal) justifies the new generation of instruments operating from the ground or from balloons (before a potential dedicated satellite mission) with unprecedented sensitivity and control of systematics.
From this perspective, we propose the QUBIC experiment, making use of the novel technique of Bolometric Interferometry, bringing together the advantages of bolometric detectors in terms of sensitivity (availability of large arrays of background-limited detectors with wide bandwidth) and of interferometry in terms of control of systematic effects (clean optics with low induced polarization, low sidelobes and therefore low ground pickup, well defined synthesized beam, small impact of individual primary beam differences) with a large number of detectors which can be replicated quite simply.

QUBIC will be composed of interferometer modules operating at three different frequencies (97, 150 and 220 GHz) with 25\% bandwidth. Each module will respectively comprise 144, 400 and 625 receiver horns whose signals will be correlated together using an optical combiner located inside each module’s cryostat. After splitting using a polarizing grid, the interference fringes will be imaged with two 900 element bolometer arrays  cooled to 100 mK in order to achieve background-limited sensitivity. The use of a rotating half-wave plate and polarized focal planes allows one to directly reconstruct the ``synthetic images" of the $I$, $Q$ and $U$ Stokes parameters observed through the primary beam.  The $TT$, $TE$, $EE$ and $BB$ power spectra can be reconstructed from the Stokes parameters' synthetic images using standard techniques. The usual calibration and a novel autocalibration technique specific to bolometric interferometry (making use of the redundancy of the array providing multiple replications of the same baselines) will allow QUBIC to achieve unprecedented control of systematics along with a sensitivity comparable to that of more traditional imaging polarimeters.
   \begin{figure*}[!ht]
   \centering\resizebox{\hsize}{!}{\centering{
   \includegraphics{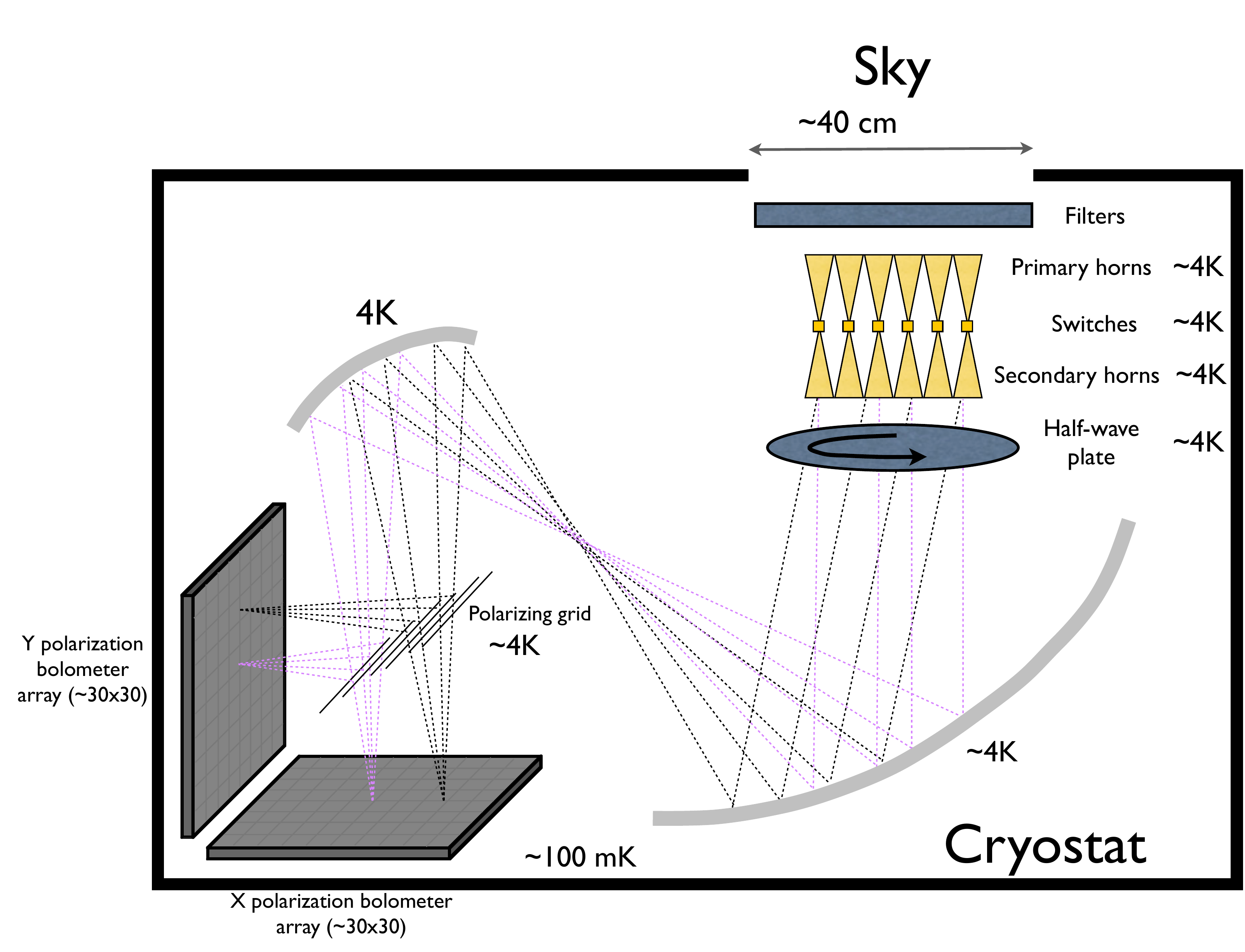}}}
   \caption{Sketch of the QUBIC concept}
              \label{sketch}
    \end{figure*}

Interferometers have a long history of successful measurements of the Cosmic Microwave Background. In fact, interferometers were the first instruments to detect the polarized anisotropies (\cite{kovac} and ~\cite{readhead}). They are generally recognized to offer certain advantages with respect to imaging systems for controlling systematic effects (insensitivity to 1/f noise, clean optics). They also offer a straightforward way to access the angular power spectrum of the signal on the sky as they provide a direct measurement of the Fourier modes of the sky observed through the primary beam.
Coherent interferometers require receivers that use either amplifiers or, for higher frequencies, SIS mixers used in a heterodyne arrangement. Any coherent receiver adds an intrinsic and irreducible amount of noise to the observed signal, preventing heterodyne interferometers from reaching background-limited sensitivity. Such interferometers are hard to scale to a large number of receiving antennas because of the complexity of the correlators, which must measure the correlations from all possible pairs of antennas.  The limited bandwidth of these systems and the requirement to measure polarization increases the complexity. 

All these reasons have led a number of teams to choose imaging instruments rather than heterodyne interferometers for the next generation of CMB polarimeters.
Bolometers are naturally wide band detectors that have negligible intrinsic noise when cooled to sub-Kelvin temperatures. They can be designed to operate at the background-limit from the ground or space. Up to now they have only been used with imagers which can now populate their focal planes with bolometer arrays with up to thousands of pixels. Imaging is a fully mature technique, widely used to observe Cosmic Microwave Background temperature and polarization anisotropies (\cite{reichardt}, ~\cite{quad}, ~\cite{takahashi}). There are however a number of sources of systematic effects that are hard to control with imagers (ground pickup and time varying systematics such as atmospheric contamination for instance) and may prevent such instruments from ultimately achieving the exquisite sensitivity required for the detection of the primordial B-modes.
Combining the advantages of interferometry in terms of systematics and direct observation of the spatial correlations of the sky on the one hand with those of imaging in terms of sensitivity in the other would obviously be of great interest for the next generation of Cosmic Microwave Background instruments (including future satellite missions). A wide band, high frequency and background-limited bolometric interferometer that could be scaled to a large number of receiving elements using large bolometers arrays would potentially be an excellent candidate for detecting such a small signal as the primordial B-modes. We describe in the present document the design we propose for such an instrument.


\section{Bolometric interferometer concept}
The bolometric interferometer we propose with the QUBIC instrument is the millimetric equivalent of the first interferometer dedicated to astronomy: the Fizeau interferometer. It was obtained by placing a mask with two holes at the entrance pupil of a telescope. Fringes were then observed at the focal plane of the telescope. In our case (see Fig.~\ref{sketch} for a sketch) we use an array of back-to-back horns acting as diffractive pupils just behind the window of a cryostat. The electric field coming from a given sky direction experiences phase differences due to the distance between the input horns. The back horns re-emit the electric field preserving this phase difference inside the cryostat.
The interference fringe patterns arising from all pairs of horns are then formed on the focal plane of an optical beam combiner which is actually just a telescope that superimposes the electric fields from all the horns at each point of the focal plane. The polarization of the incoming field is modulated  using a half-wave plate located after the back horns. A polarizing grid separates both polarizations towards two different focal planes each measuring a linear combination of $I$, $Q$ and $U$ Stokes parameters.

The fact that the electric fields from all horns are added and then squared and averaged in time using the bolometers makes our instrument an adding interferometer (Fizeau combination), in contrast to radio-interferometers, where the signals (visibilities) are obtained by multiplying the electric fields from pairs of receivers  (Michelson combination) using an analog or digital correlator. We do not need a large number of complex and expensive correlators as the correlation between channels is naturally achieved with the bolometers: $\left< \left| E_1+E_2\right|^2\right> = \left< \left| E_1\right|^2\right>+\left< \left| E_2\right|^2\right>+2\left< \mathrm{Re}(E_1 E_2^\star)\right>$.   The first sum,  $\left< \left| E_1\right|^2\right>+\left< \left| E_2\right|^2\right>$, which is just the power in the primary beam, is just a background power\footnote{Note that this DC term will vary with time if we scan the sky, but on very large timescales due the wide primary beam.}. The last term, the interference term, is proportional to the visibility.  This simple beam combiner allows the instrument to be scalable to a large number of input horns.

We plan to install a first QUBIC module at the Franco-Italian Concordia station in Dôme C, Antarctica within two years. The first module will consist of an array of 400 horns operating at 150 GHz with 25\% bandwidth and 14 degree (FWHM) primary beams. The optical combiner will have a focal length of $\sim 30$ cm and each of the two focal planes will be comprised of arrays of 30x30 bare TES bolometers of 3 mm size. The full instrument will include modules at three different frequencies (90, 150, 220 GHz) and will constrain a tensor to scalar ratio of 0.01 in one year of data taking at the 90\% confidence level.

\section{Instrument sub-systems}
We describe in this section the various sub-systems of the QUBIC instrument. Most of them are already available within the collaboration but a few still require a reasonable amount of R\&D (ongoing within the collaboration) to be fully operational. Note that all components required for QUBIC are similar to those required for other CMB polarization experiments based on the traditional imager concept.

\subsection{Cryostat and cryogenics}
The first cryogenic stage of the cryostat  will be provided by a pulse-tube cooler that allows continuous operation of the instrument at a temperature below 4K. It will cool the filters, horns, half-wave plate, optical combiner and polarizing grid. The second cryogenic stage supporting the detectors at 100mK will be provided by a dilution refrigerator. The large ($\sim$40 cm) window on top of the cryostat will use Zotefoam to provide good mechanical strength and low emissivity.

\subsection{Cold optics}
\subsubsection*{Horns}
The requirements for the front-end (primary) horn antennas of QUBIC are similar to those of previous and current CMB imaging experiments. The antennas should have low return loss and have well-understood beams characterized by low cross-polarization, and low sidelobe levels. They should have good far-field beam circularity and operate over bandwidths up to about 30\%. Because of these stringent optical performance requirements, single-mode corrugated horns have to date been favored in CMB experiments. This type of horn has already been shown to meet the exacting requirements listed above and is readily manufacturable at the QUBIC observing frequencies. Single-mode horns allow only one orthogonal pair of coherent fields to propagate and so have well-understood polarization characteristics. Few-moded horns, while they have the advantage of increasing throughput (\cite{murphy2010}), are likely to scramble polarization information and may not meet the beam circularity or sidelobe criteria so easily.

The length of the horns (if, for example, weight needs to be minimized) and the sidelobe levels can be reduced by adjusting the horn profile away from the standard conical shape (e.g. ~\cite{osullivan}, \cite{maffei}). Shaping can also be used to increase the aperture efficiency of a horn by moving the phase centre close to the aperture (as could a phase-flattening lens). Careful design of waveguide transitions keeps the return loss very low. We are considering either of the two types of horns developed for the Clover project. At 150 GHz ``Ultra-Gaussian" horns (so- called because 99.9\% of power is in the free-space fundamental Gaussian mode) were designed (Ade et al., 2009). In these horns the $\rm HE_{12}$ mode is deliberately excited in a cosine-squared profile section of the horn and then brought into phase with the dominant $\rm HE_{11}$ mode in an extended parallel front section. At 97 GHz corrugated Winston-like profiled horns (\cite{maffei2004}) were used and shown to produce very low spillover (\cite{grimes2009}).

\subsubsection*{Filters}
Spectral filters will be located inside the dewar to perform several roles. First, the selection of the spectral band of observation is achieved by the combination of the back-to-back waveguide section (frequency cut-on) and metal-mesh interference filters (frequency cut-off).  Second, metal mesh interference filters located at the entrance aperture of each thermal stage reduce the radiation load on the cryogenic system and the detectors.  The radiation load reaching the detectors is not only a source of stray-light but will also lead to a poorer sensitivity if not well under control.

\begin{itemize}
\item {\it 300K stage:} The dewar window will already act as a filter by reflecting and absorbing most of the short wavelength radiation, up to the near-infrared. 
\item {\it Intermediate thermal stage filters in front of the back-to-back horns:}
Each screen aperture will have at least one low-pass edge filter with possibly an associated thermal (IR) filter to reduce the thermal loading. These filters and the 300K stage window and filter, will all have to be carefully designed and characterized. Being located in front of the back-to-back horn apertures, they can affect the beam shape of the instrument. Previous measurements have shown that the best components have a minimal impact on the beam shape and an induced instrumental polarization lower than -35dB. However they will have to be accurately characterized, typically down to -50dB.
In order to clear the main beam of all the horns, and taking into account the number of pixels per instrument, a clear aperture diameter of 400mm is required for all these filters and windows.
\item {\it Filters located after the back-to-back horns:}
Probably two thermal stage apertures will be between the back-to-back horns and the detectors (4K and ~100mK). Although the filters located at these apertures will be manufactured with the same attention as the previous ones, because the polarization will have been already modulated, they will have a lesser impact on systematic effects. The focus will be on achieving a high out-of-band rejection in order reduce stray-light. The spectral band defining filter will be located on the aperture of the box shielding the detectors in order to be as close as possible to these.
\end{itemize}

\subsubsection*{Switches}
The switches are required during the calibration procedure but are not used during data acquisition. They are used as shutters that are operated independently for all channels. The idea is to modulate a given visibility measured by a given pair of horns at one time in order to compare it with other pairs of horns measuring the same (equivalent or redundant) baseline. This has proven to be an extremely powerful self-calibration tool in interferometry (see section~\ref{syste} for details). The switches will metal pins or films that close the waveguide section between the back-to-back horns. They will be operated using miniature electromagnets commanded from the outside of the cryostat. The switching does not need to be fast but will need to be reliable.

\subsubsection*{Half-wave plate}
The rotating half-wave plate modulates the signal coming from the sky. It is therefore a critical component that needs to be designed carefully and tested extensively. The systematic effects that could be generated by the half-wave plate are reduced when cooled to low temperature (\cite{salatino1}). The inhomogenities of the half-wave plate are however an issue for such large modulator. Possible solutions to manufacturing a broadband half-wave plate include either birefringent plates like sapphire for example (\cite{hwp1}) or a stack of metal mesh filters (\cite{hwp2}). 
The rotating mechanism will be based on the design made for the PILOT experiment (\cite{salatino1}), allowing for either a continuous rotation speed or a step rotation with a precision better than 0.1 deg.

\subsubsection*{Optical combiner}
Signal combination in QUBIC will be performed by means of an optical system that transfers the fields radiated by the primary back-to-back horn array to a detector plane where Fizeau interference fringes can be observed. The QUBIC system must satisfy the following basic requirements:
\begin{itemize}
\item The combiner is an optical system in which rays launched at a given angle from the re-emitting (back) horn array are focused to a single point on the focal plane. In this way (for an ideal optical system) equivalent baselines will produce identical fringe patterns.
\item The limit on the total number of bolometers that can be produced, together with the sampling requirement for at least two bolometers per fringe, constrains the equivalent focal length of the combiner to range from 200 to 300 mm. Since the back-to-back horn array size is ~240 mm in diameter for the 150 GHz instrument, the combiner will be a very fast system (small F\#).
\item We can use mirrors or lenses as optical components. Our choice will be between an on-axis lens system or an off-axis mirror system that can avoid the shadowing of any baselines (on-axis mirrors would result in very high levels of truncation). On one hand the behavior of mirrors can be extremely well characterized using  physical optics, but on the other an off-axis system will introduce aberrations.
\item The system has to be compact enough to be placed in a cryostat of about 1m$^3$.
\end{itemize}
The collaboration is investigating different layouts for the combiner, possibly including lenses, starting with classical astronomical telescope designs such as Gregorian and Cassegrain, and bearing in mind the constraints above. We find that CATR designs (Compact Test Range, crossed Dragone configuration -- \cite{tran}), which have proven suitable elsewhere, are not quite feasible given our very short focal length and wide field-of-view. We are therefore currently studying the optimization of an off-axis Gregorian telescope. 


\subsection{Detector chain}
The focal plane will be covered with a polarized filled array in order to sample properly the image. The simplest way to separate polarization is to use a polarizing grid which directs the beam to two detector arrays, one in transmission and one in reflection. Taking into account the constraints from the optical combiner and the size of the secondary horns leads to the need for about 900 detectors of 3mm size each for the 150 GHz module. The main performance requirements for the detectors are their NEP and time constant. With a background power of few pW, the detector NEP needs to be lower than about $5-10\mathrm{aW.Hz}^{-1/2}$ at 150GHz. The modulation speed of the half wave plate of a few Hz constrains the detector time constant to be shorter than 10ms. These requirements are easily reached with 100mK TESs which are the current baseline. 

\begin{figure}[!t]
   \centering\resizebox{\hsize}{!}{\centering{
   \includegraphics{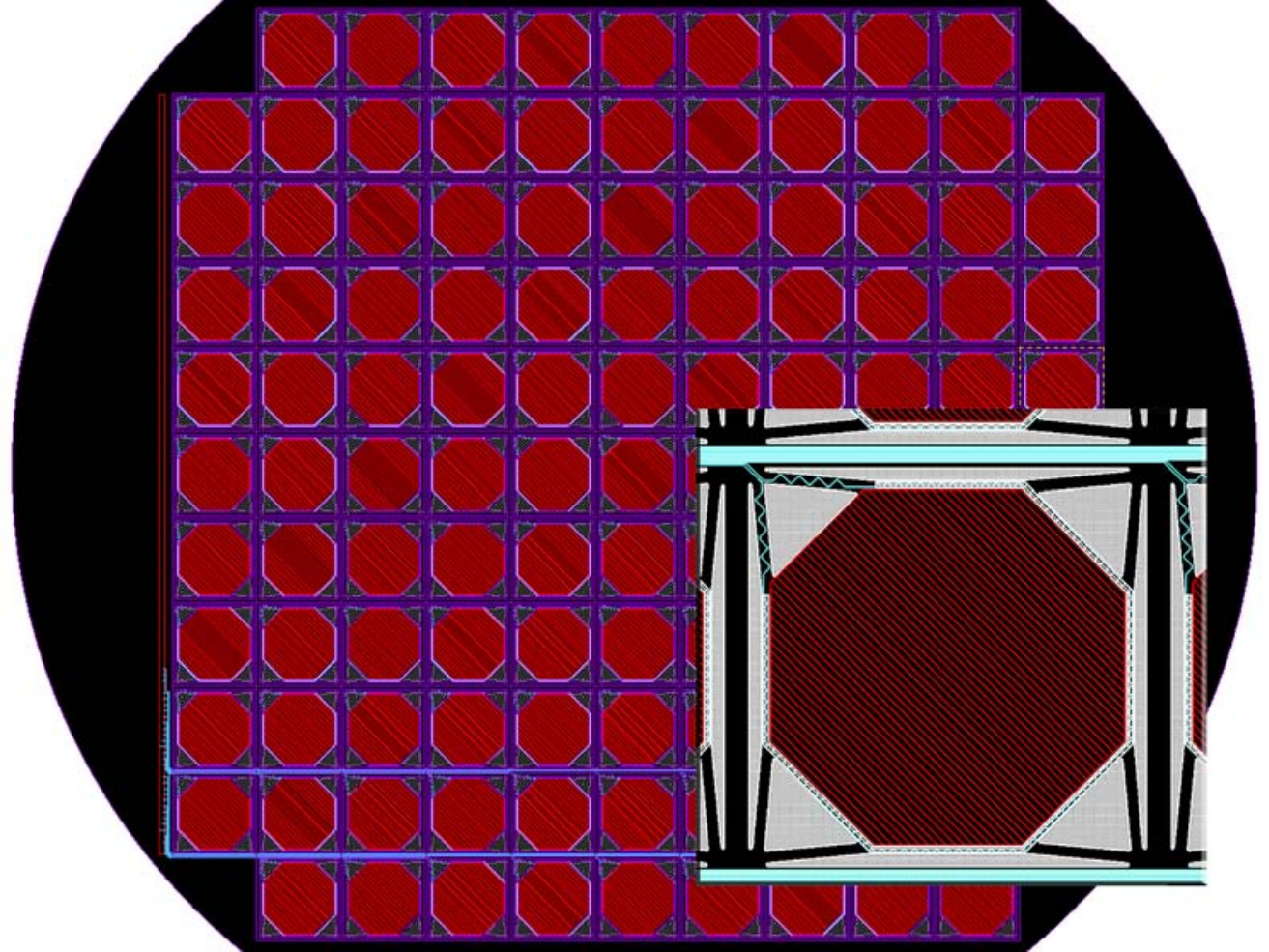}}}
   \caption{Mask design for a 128 elements TES array on a 2 inch silicon wafer. We have also displayed a zoom of one of the detectors on the bottom-right.}
              \label{tes}%
    \end{figure}

The detector assembly sub-system will be based on NbSi transition edge sensors (TES) read out with a time domain multiplexing scheme (\cite{TES1,TES2}). While other technologies like KIDs could be used, TES array technology is currently more fully developed. The use of NbSi alloy as the thermal sensor offers several advantages with respect to classical bi-layers: the sensors are more homogeneous on a large array, the thermal response is high and the  noise properties of NbSi alloy is intrinsically better than bilayers. A first design of a detector array for the focal plane is shown in Fig.~\ref{tes}.

The TES readout electronics use SQUIDs as a first amplifier stage followed by a 4K SiGe ASIC that controls the multiplexing and amplifies the signal from the SQUIDs (\cite{TES3,TES4}). The advantages of such a system are a simplification of the architecture, miniaturization of the readout electronics and also immunity to high energy particles along with an overall low power consumption and very low noise properties. We have demonstrated this technique with a 24 pixel readout in a 24:1 multiplexing scheme (done in two steps: 8:1 by SQUIDs stage and 3:1 with the ASIC). The current design is based on a new ASIC under development that will be able to readout 128 detectors with a 128:1 multiplexing factor (32:1 SQUIDs and 4:1 ASIC). The full detector array will be therefore read out with 7 of these ASICs.

\subsection*{Mount}
The instrument will be installed on a classical alt-azimuthal mount allowing for three axis motion in order to both scan in azimuth and elevation and to rotate the instrument around its optical axis.

\section{Data analysis for bolometric interferometry}
\subsection{Synthetic image}
One can express the bolometer measurements as a linear combination of the visibilities (basically Fourier modes) of the sky observed through the primary beam of the horns. The coefficients of the linear combination of visibilities arriving at each bolometer are just related to the phase-differences between different channels imposed by the optical combiner. This was explored in detail in~(\cite{muxpaper})  where we have shown that the visibility reconstruction can be done in an optimal way. Working with the visibilities, although it may seem logical as we are dealing with an interferometer, may not be the best approach however. It is better to use our instrument as a {\em synthetic imager}.
As will be shown in this section, the image in our focal plane is  exactly the quantity usually known in interferometry as the {\em synthetic image} and is usually obtained through an inverse Fourier transform of the visibilities. In our case, the visibilities are not the natural observable (in contrast with a multiplicative interferometer) as we have immediate access to the synthetic image. We can however recover the visibilities through a Fourier transform of the synthetic image. The synthesized beam we convolve with the sky is determined by the configuration of the primary horns, its resolution being basically given by the largest separation between horns. Once this synthesized beam is known, one can use our instrument as a simple synthetic imager by scanning the sky as with an usual imager, making maps and calculating power spectra using standard techniques. It is however completely an interferometer in the sense that the synthetic beam is well defined by the location of the primary horns. One of the most important advantages of having an interferometer is the fact that, using the redundancy of the baselines, one can internally calibrate the instrument accurately, including systematic effects such as gains and cross-polarization.  This topic will be discussed in section~\ref{syste}.

The signal on the bolometers as a function of time is\footnote{This is however an approximate expression as multiple reflexions will modify it, producing terms at $\omega$ and $3\omega$ that have to be dealt with (\cite{salatino1}, \cite{salatino2}).}:
\begin{equation}
R(\vec{d}_p, t) = S_I(\vec{d}_p)\pm \cos(4\omega t)S_Q(\vec{d}_p)\pm \sin(4\omega t)S_U(\vec{d}_p)\label{signal}
\end{equation}
where the $\pm$ is $+$ for one of the focal planes (polarized in one direction)  and $-$ for the other one polarized in the other direction. $\vec{d}_p$ is the location of the detector in one of the focal planes and $\omega$ is the rotation frequency of the half-wave plate.  The combination of hase-sensitive detection at the half-wave plate modulation frequency and scanning across the sky allows recovery of the three focal plane images, $S_I(\vec{d}_p)$, $S_Q(\vec{d}_p)$ and $S_U(\vec{d}_p)$, corresponding to the three Stokes parameters.

For the $X$ Stokes parameter the synthetic image can be directly expressed as a function of the observed sky:
\begin{equation}
S_X(\vec{d}_p)=\int X(\vec{n})B^p_s(\vec{n}) \mathrm{d}\vec{n}
\end{equation}
where $B_s^p$ is the {\em synthesized beam} of the interferometer for detector $p$ that is formed by the arrangement of the array of input horns. 

\subsection{Synthesized beam}
The synthesized  beam is just the sum of all wide band fringe patterns formed by all pairs of horns modulated by the primary beam on the sky and by the secondary beam on the focal and integrated over the bolometer surface:
\begin{eqnarray}
B_s^p(\vec{n})&=&B_\mathrm{prim}(\vec{n}) \int\int   B_\mathrm{sec}(\vec{d}) \label{eqsynthbeam} \\
&&~~~~~\times \left | \sum_i \exp\left[ i2\pi \frac{\vec{x}_i}{\lambda} \cdot \left( \frac{d}{D_f}-\vec{n}\right)\right]\right|^2 J(\vec{\nu})\Theta(\vec{d}-\vec{d}_p) \mathrm{d}\nu \mathrm{d}\vec{d} \nonumber
\end{eqnarray}
where $D_f$ is the equivalent focal length of the optical combiner, $B_\mathrm{prim}$ is the primary beam on the sky of the input horns, $B_\mathrm{sec}(\vec{d})$ is the beam of the reemitting horns as seen on the focal plane ($\vec{d}$ labels the position on the focal plane). The integration is performed over the bandwidth of the instrument $J(\nu)$ ($\nu$ is the frequency and $\lambda=c/\nu$ the wavelength) and over the surface of the bolometer indexed by $p$ and modeled with the top-hat like function $\Theta(\vec{d}-\vec{d}_p)$. The sum in the integral is performed over the input horns with position given by $\vec{x}_i$. We have assumed in the above expression that all horns have the same primary and secondary beams so that they can be factorized. A more realistic case can be easily accounted for in the calculation of the primary beam, provided the fact that individual beams are well known (from calibration on point sources for instance).
   \begin{figure}[!t]
   \centering\resizebox{\hsize}{!}{\centering{
   \includegraphics{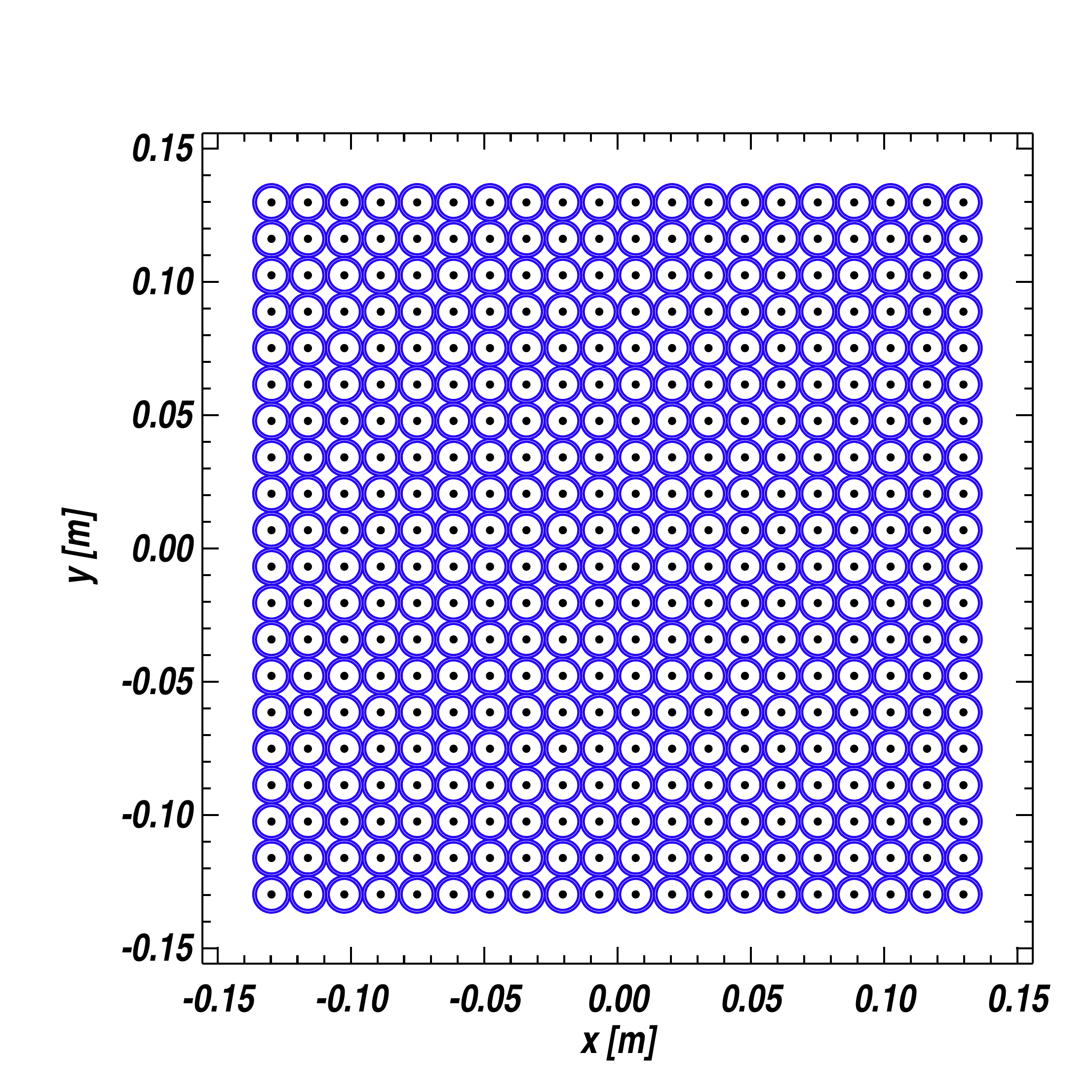}}}
   \caption{20x20 horn array configuration for the 150 GHz QUBIC module. Each horn has a beam of 14 degree (FWHM), has an internal radius of 5.8 mm, and has 1 mm thick walls.}
              \label{array}
   \centering\resizebox{\hsize}{!}{\centering{
   \includegraphics{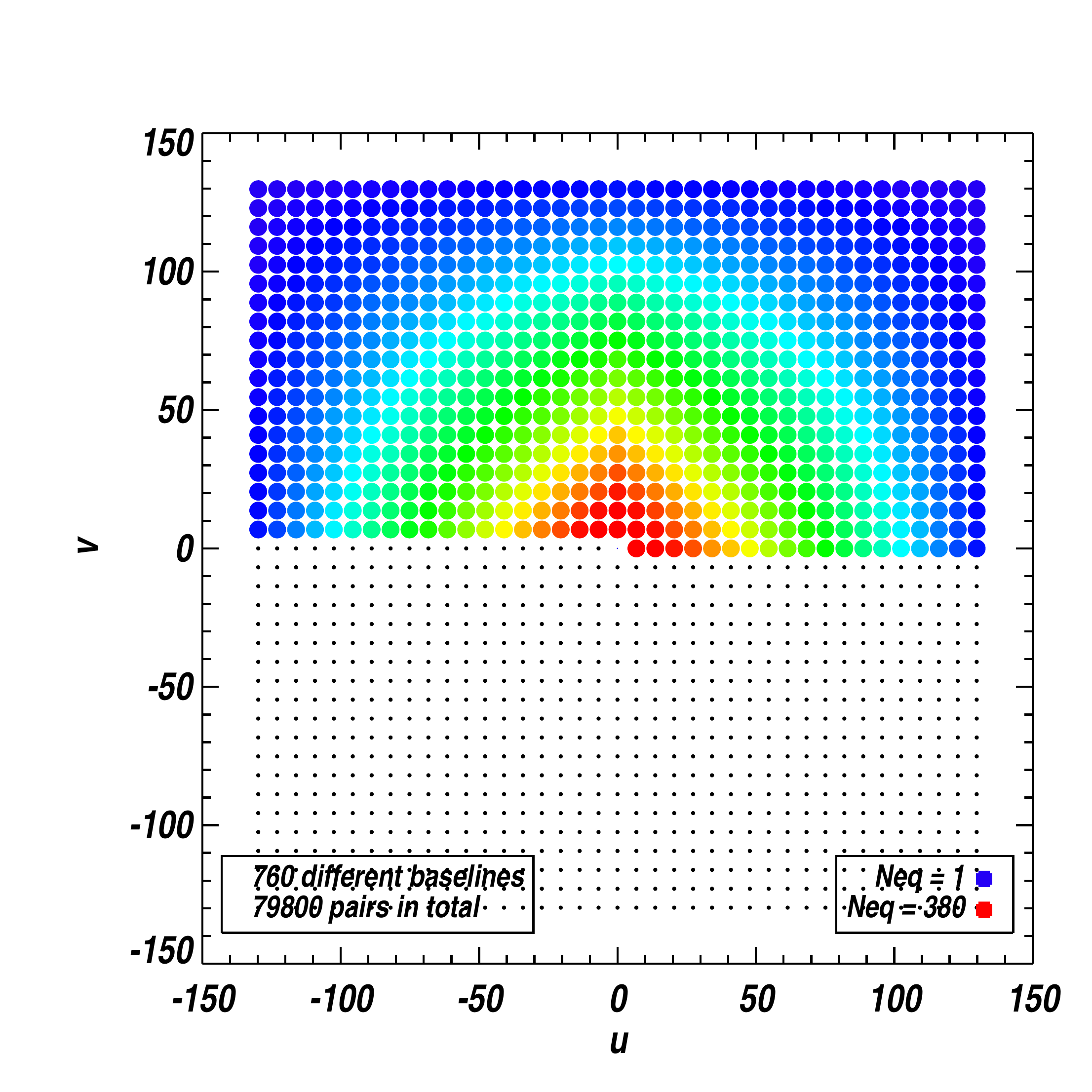}}}
   \caption{{\em uv-plane} coverage of the array shown in Fig.~\ref{array}. Each point corresponds to a different baseline and is highly redundant in our array. The number of equivalent baselines is shown in colors. The multipoles can be calculated from the baseline length by multiplying by $2\pi$ (this is exact in the flat sky approximation, but we are not in this range, so the correspondence is approximate).}
              \label{redundancy}
    \end{figure}

If one had an infinite number of primary horns, the synthetic beam would be a Dirac peak in the center of the field of view of the primary beam. We have chosen to place the primary horns on a compact square array as shown in Fig.~\ref{array} in order to achieve a maximal redundancy of the baselines (see Fig.~\ref{redundancy}) defined by each possible pair of horns. Such a redundancy was shown to be crucial to ensure an optimal sensitivity of the bolometric interferometer (\cite{muxpaper}) allowing for what we have called a ``coherent summation of equivalent baselines". This means that all equivalent (or redundant) baselines need to experience the same phase difference configuration on the focal plane in order to maximize the signal to noise ratio of the visibilities, or of the synthetic images. Such a condition can be easily understood: on the focla plane, each baseline contributes a sinusoidal fringe pattern with a spatial frequency determined by the baseline length. A second baseline equivalent to the first one would contribute with the same pattern shifted if both baselines were not ``phase-locked". With a large number of equivalent baselines $N_\mathrm{eq}$, the fringe signal would be washed out as $1/\sqrt{N_\mathrm{eq}}$ while if all baselines are coherent in phase, the fringe signal is just summed $N_\mathrm{eq}$ times without being washed out. The use of an optical combiner automatically ensures that equivalent baselines are summed coherently (\cite{muxpaper}). A compact square array of primary horns ensures a large number of equivalent baselines (hence a good sensitivity) while allowing for an almost maximal density of horns within the same aperture window.

With such a redundant array, in the monochromatic and point-like detector limit, the integrals in Eq.~\ref{eqsynthbeam} disappear and the synthetic beam can be analytically calculated:
\begin{eqnarray}
B_s^p(\vec{n})&=&B_\mathrm{prim}(\vec{n}) B_\mathrm{sec}(\vec{d}_p) \nonumber\\
&&~~~~~\times \frac{\sin^2\left[P \pi \frac{\Delta x}{\lambda}\left( \frac{d_{xp}}{D_f}-n_x\right)     \right]}{\sin^2\left[\pi \frac{\Delta x}{\lambda}\left( \frac{d_{xp}}{D_f}-n_x\right)     \right]}
\frac{\sin^2\left[P \pi \frac{\Delta x}{\lambda}\left( \frac{d_{yp}}{D_f}-n_y\right)     \right]}{\sin^2\left[\pi \frac{\Delta x}{\lambda}\left( \frac{d_{yp}}{D_f}-n_y\right)     \right]}
\end{eqnarray}
where $P=20$ is the number of horns on a side of the square array, $\Delta x$ is the distance between two horns  and the indices $x$ and $y$ denote projection along the respective axis in the focal plane reference frame. $\ell_\mathrm{min}\sim 2\pi\frac{\Delta x}{\lambda} \sim 43$ is the minimum multipole accessible with this array and a single field observation while $\ell_\mathrm{max}\sim \sim 2\pi\frac{P\Delta x}{\lambda} \sim 867$ is the largest multipole accessible (although with a small number of equivalent baselines, and hence a very poor sensitivity - the actual maximum for well-measured multipoles is around 200).
The maximum multipole accessible defines the angular resolution of the synthetic beam to be $\sim \frac{2\pi}{\ell_\mathrm{max}}\sim 0.4$ degrees while the fact that the minimum multipole accessible is $\ell_\mathrm{min}$ implies that this small synthetic beam replicates over the sky with peaks separated by $\sim \frac{2\pi}{\ell_\mathrm{min}}\sim 8.5$ degrees. This replication is however significantly reduced by the apodization of the primary beam that defines the field of view of our interferometer (we have chosen 14 deg FWHM). The peaks of the synthetic beam are shifted relative to the center of the field of view for different detector locations in the focal plane. The primary beam however remains fixed so that the relative intensity of the synthetic beam peaks changes from one detector to another.

\begin{figure}[!t]
   \centering\resizebox{\hsize}{!}{\centering{
   \includegraphics{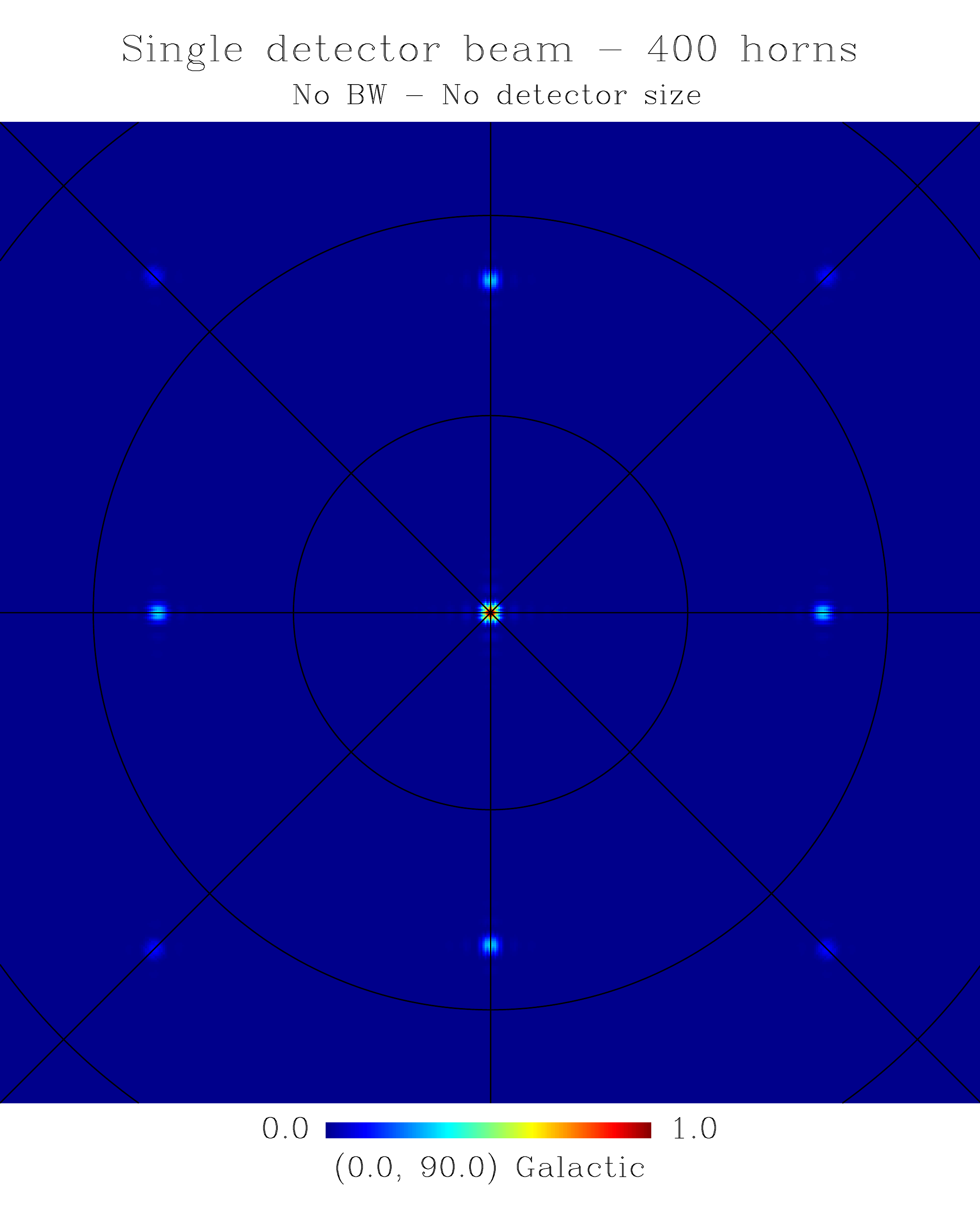}~\includegraphics{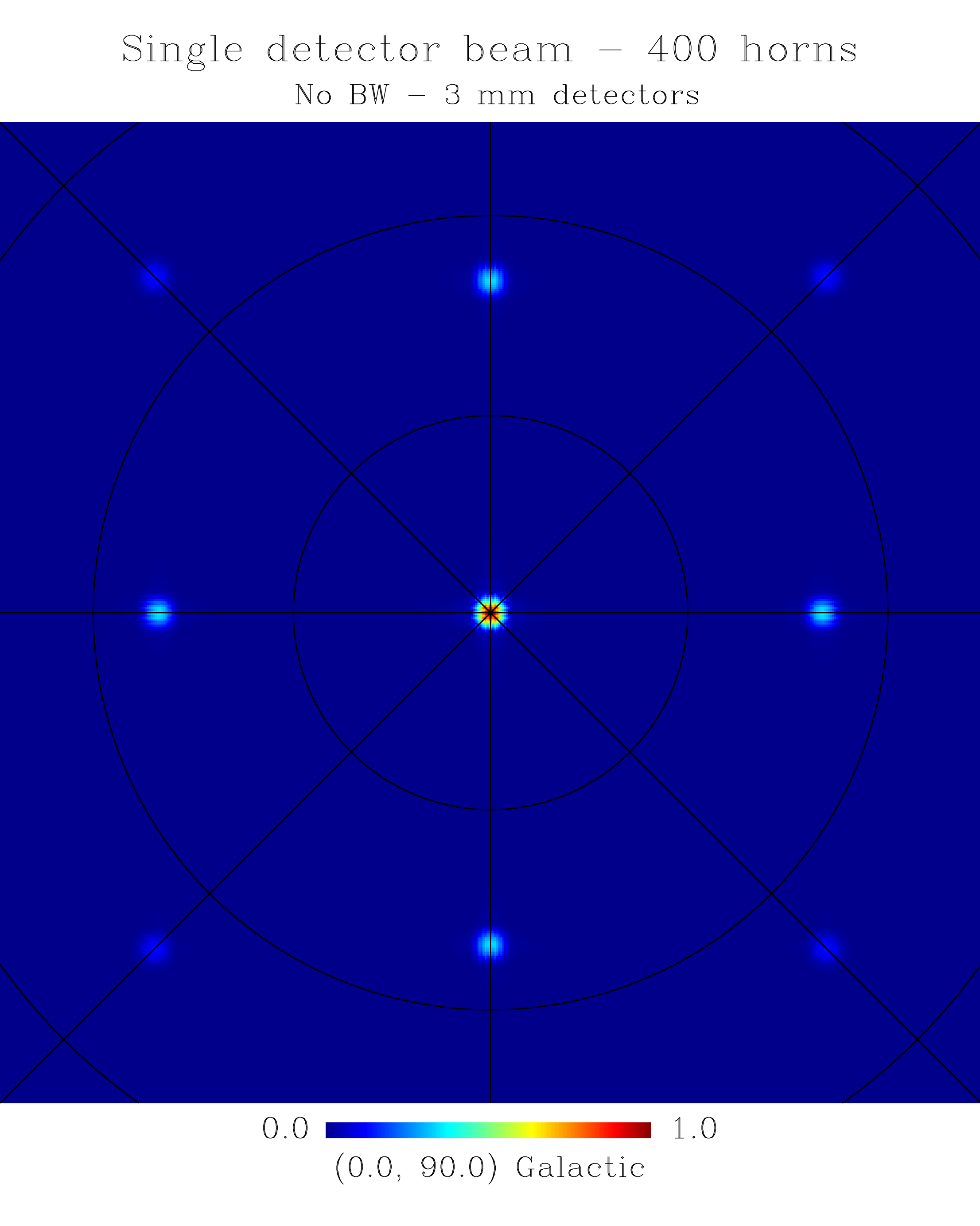}}}
   \centering\resizebox{\hsize}{!}{\centering{
   \includegraphics{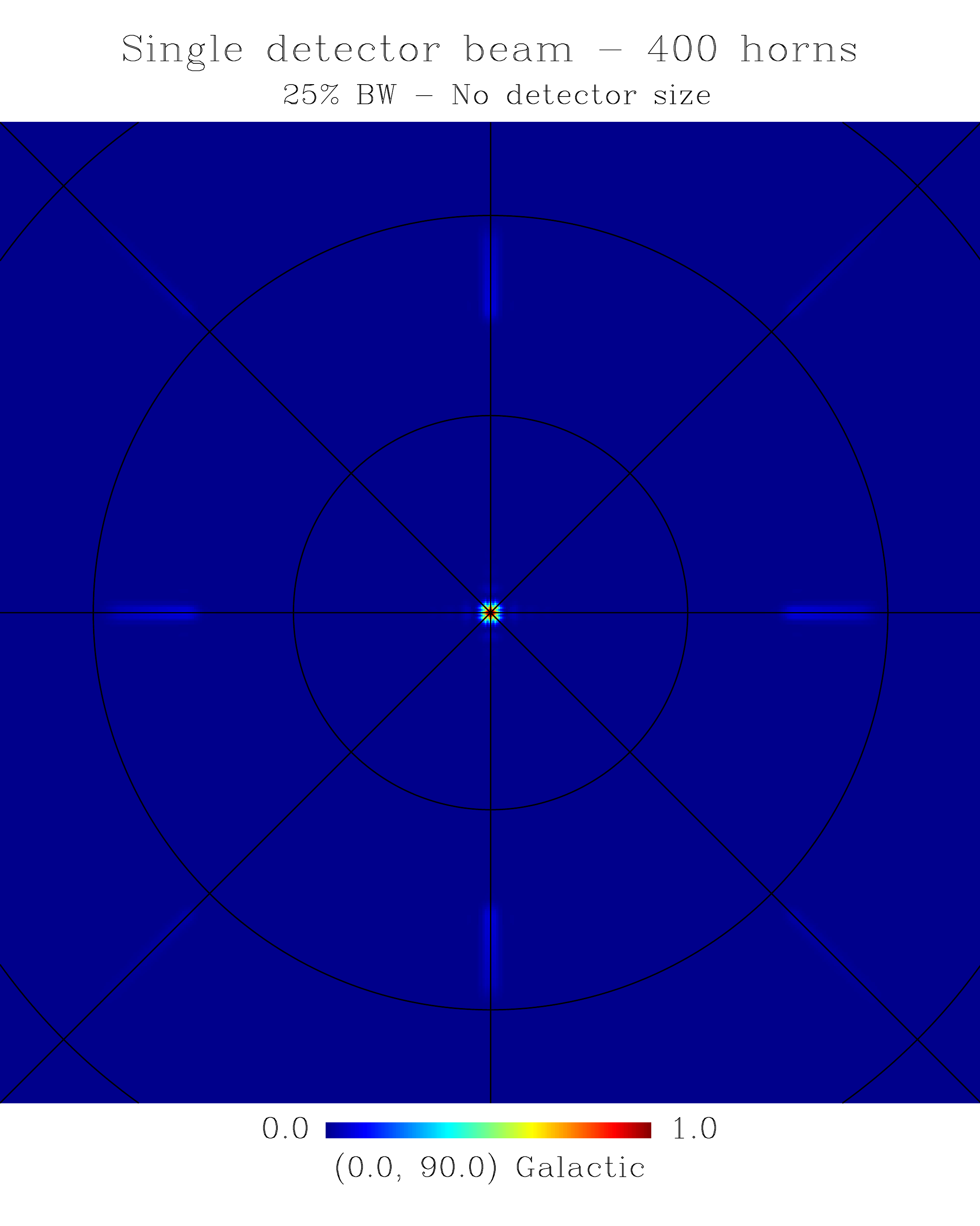}~\includegraphics{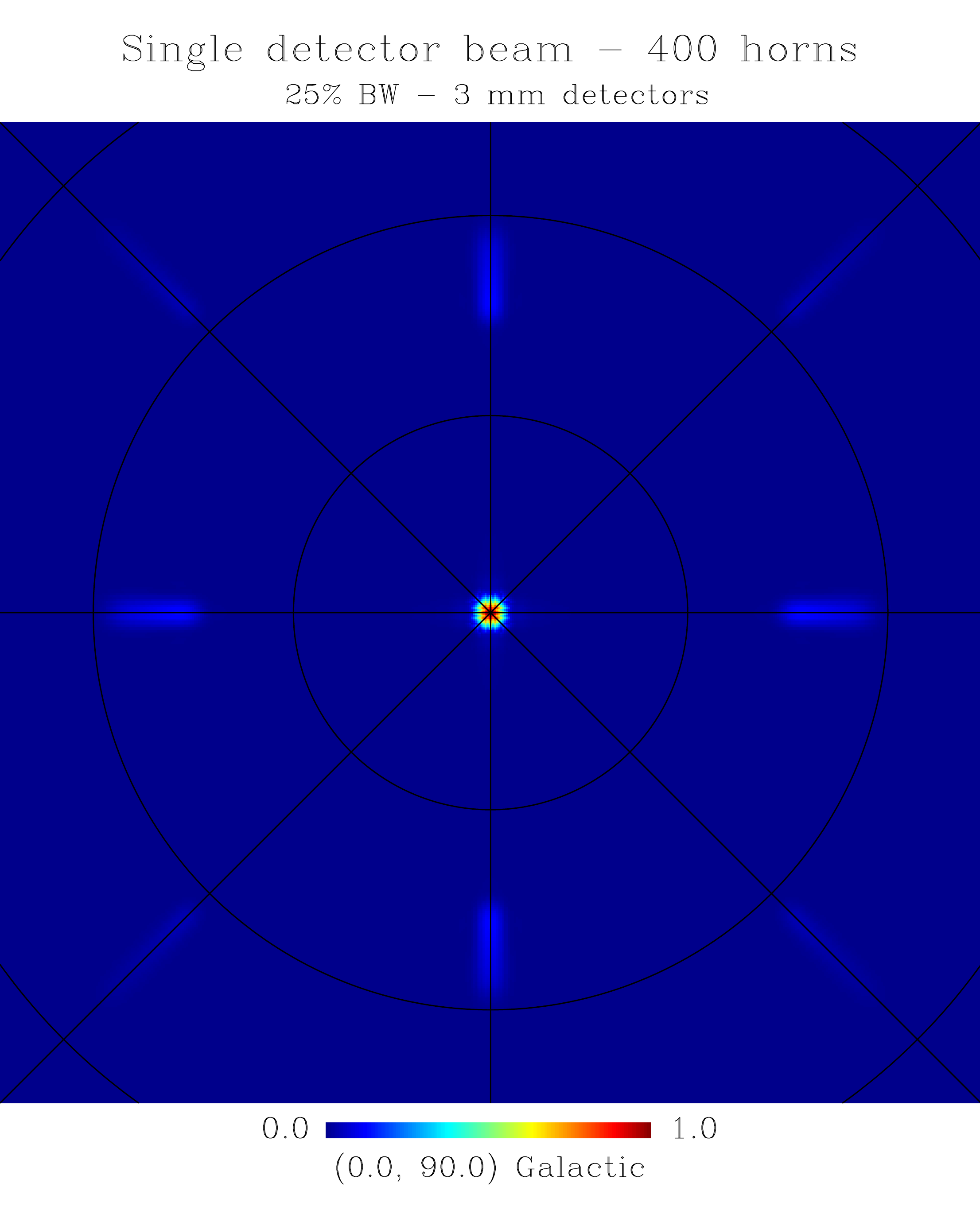}}}
   \caption{Gnomonic projections of Healpix (\cite{healpix}) maps of the QUBIC synthesized beam for a monochromatic instrument with point-like (top-left) or 3 mm detectors (top-right), a 25\% bandwidth instrument with point-like (bottom-left) and 3 mm detectors (bottom-right). All images were obtained with a compact square 400 horns array with 14 degrees FWHM beams at 150 GHz. The effect of the integration over bandwidth and detectors can easily be seen, the former only affects the shape far from the center while the latter slightly enlarges the peaks. The black circles are at 5 and 10 degrees from the center.}
              \label{synthbeam}%
    \end{figure}

\begin{figure}[!h]
   \centering\resizebox{\hsize}{!}{\centering{
   \includegraphics{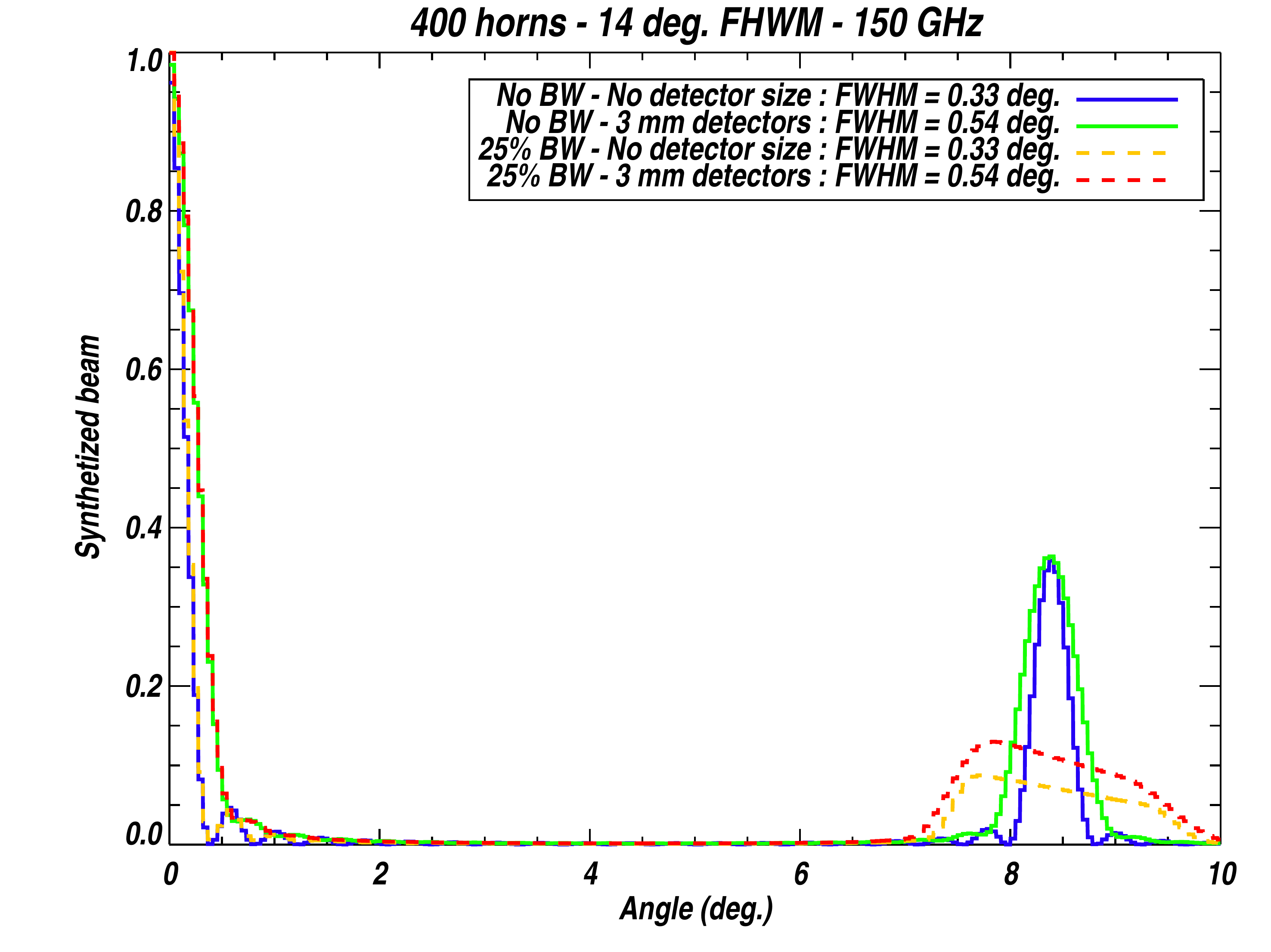}}}
   \caption{Horizonthal cut of the synthesized beam maps shown in Fig.~\ref{synthbeam}. The FWHM of the central peak is obtained through a Gaussian fit to the profile. One sees the slight loss of resolution due to the finite detector size while the effect of bandwidth (dashed lines) can be seen on the replicated peaks, far from the center of the field of view. The next peak not displayed in the figure (at $\sim 17$ degrees from the center) peaks at $\sim 0.015$ for the blue and green lines and $\sim 0.006$ for the orange and red lines. The following one (at an angle of $\sim$25.5 deg) peaks around $\sim 5\times 10^{-5}$. These numbers assume a perfect gaussian primary beam and might be larger in the presence of sidelobes in the primary beams.}
              \label{synthbeamcut}%
    \end{figure}

The synthetic beam is shown in Fig.~\ref{synthbeam} for monochromatic or broadband cases and point-like or finite-size detectors (located at the center of the focal plane) showing the effect of the integration over bandwidth and detector surface. We have used an array of 400 primary horns with 14 degree (FWHM) primary beam and a combiner with an equivalent focal length of 30 cm, a central frequency of 150 GHz and 25\% bandwidth. We also show in Fig.~\ref{synthbeamcut} horizontal cuts of the maps in Fig.~\ref{synthbeam} showing the profiles of our synthesized beam. We see that integration over the pixel size slightly degrades the resolution of the central peak from 0.33 degree to 0.54 while integration over bandwidth only affects signal far from the center of the field of view. We also note that as expected, the replicated peaks are separated by $\sim 8.5$ degrees.

\subsection{Power spectrum from the synthetic image}
In the previous section, we have shown that the direct observable of a bolometric interferometer such as QUBIC is the synthetic image, usually obtained in traditional interferometry from the inverse Fourier Transform of the visibilities directly measured by the set of horns and correlators. Our bolometric interferometer is therefore very different from other interferometers in this respect and somehow closer to an imager whose instantaneous observable is also the sky convolved with the instrumental beam\footnote{In an imager, this beam is formed by the antenna that couples to each detector projected on the sky through the telescope (reducing the horn beam from a few degrees to a few arc-minutes depending on the telescope diameter).}. 

The visibilities in an interferometer are quantities that are defined as the Fourier transform of the observed field projected on the plane tangent to the observing direction. These are just the modes on the sky if the primary is small enough so that the {\em flat sky} approximation is valid. With a 14 degree primary field of view, this approximation is clearly not valid. It is therefore not straightforward to combine visibilities taken while observing different directions of the sky with a wide field of view interferometer (although it has been explored in detail in~\cite{mosaicking}). With our synthetic imager approach, the problem is naturally solved:
as the synthetic image is simply the convolution of the sky through the synthetic beam, one can directly relate the covariance matrix of the bolometer data to the angular power spectrum. Let's introduce $\beta_{\ell m}(p,\vec{n}_0, \theta_0)$:
\begin{equation}
\beta_{\ell m}(p,\vec{n_0},\theta_0)=\int B_s^p(\vec{n}-\vec{n}_0,\theta_0)Y_{\ell m}(\vec{n}) \mathrm{d}\vec{n}
\end{equation}
where $B_s^p(\vec{n}-\vec{n}_0,\theta_0)$ is the synthesized beam for detector number $p$ in the focal plane when the intrument points towards $\vec{n}_0$ with a pitch angle $\theta_0$ (rotation around the $\vec{n}_0$ direction)\footnote{Note that $\beta_{\ell m}(p,\vec{n_0},\theta_0)$ is not the spherical harmonic coefficient of $B_s^p(\vec{n}-\vec{n_0},\theta_0)$ (which would have required a $Y_{\ell m}^\star$), but the conjugate of the spherical harmonics transform of $B_s^p~^\star(\vec{n}-\vec{n_0},\theta_0)$.}.

It is straightforward to show that the total intensity synthetic image (measured with bolometer $p$ when pointing towards $\vec{n}_0$ with pitch angle $\theta_0$) is:
\begin{equation}
S_I(p,\vec{n}_0,\theta_0)=\sum_{\ell m} a_{\ell m} \beta_{\ell m}(p,\vec{n_0},\theta_0)
\end{equation}
where the $a_{\ell m}$ are the underlying sky spherical harmonic expansion coefficients. The covariance matrix of the bolometer signals can be explicitly written as the following:
\begin{equation}
\left< S_I(p,\vec{n_k},\theta_i) \cdot S_I^\star (p',\vec{n_l},\theta_j)\right> = \sum_\ell C_\ell W_{\ell} (p,p',\vec{n_k},\vec{n_l},\theta_i,\theta_j)
\end{equation}
where we have introduced the window function:
\begin{equation}
W_{\ell} (p,p',\vec{n_k},\vec{n_l},\theta_i,\theta_j) = \sum_m  \beta_{\ell m}(p,\vec{n_k},\theta_i) \beta^\star_{\ell m}(p',\vec{n_l},\theta_j) \label{eqwl}
\end{equation}
Similar window functions can be calculated in a straightforward manner for polarization of course, so that the covariance matrix of the synthetic images of $I$, $Q$ and $U$ can be expressed as a function of these window functions and the power spectra $C^{TT}$, $C^{TE}$, $C^{EE}$, $C^{BB}$ (as well as $C^{TB}$ and $C^{EB}$ generally required for consistency checks). If one observes a limited number of fields with a limited number of pitch angles so that the number of bolometer data samples involved is not too large, the power spectrum can be estimated using a brute force maximum likelihood using the window functions in Eq.~\ref{eqwl}.

   \begin{figure}[!t]
   \centering\resizebox{\hsize}{!}{\centering{
   \includegraphics{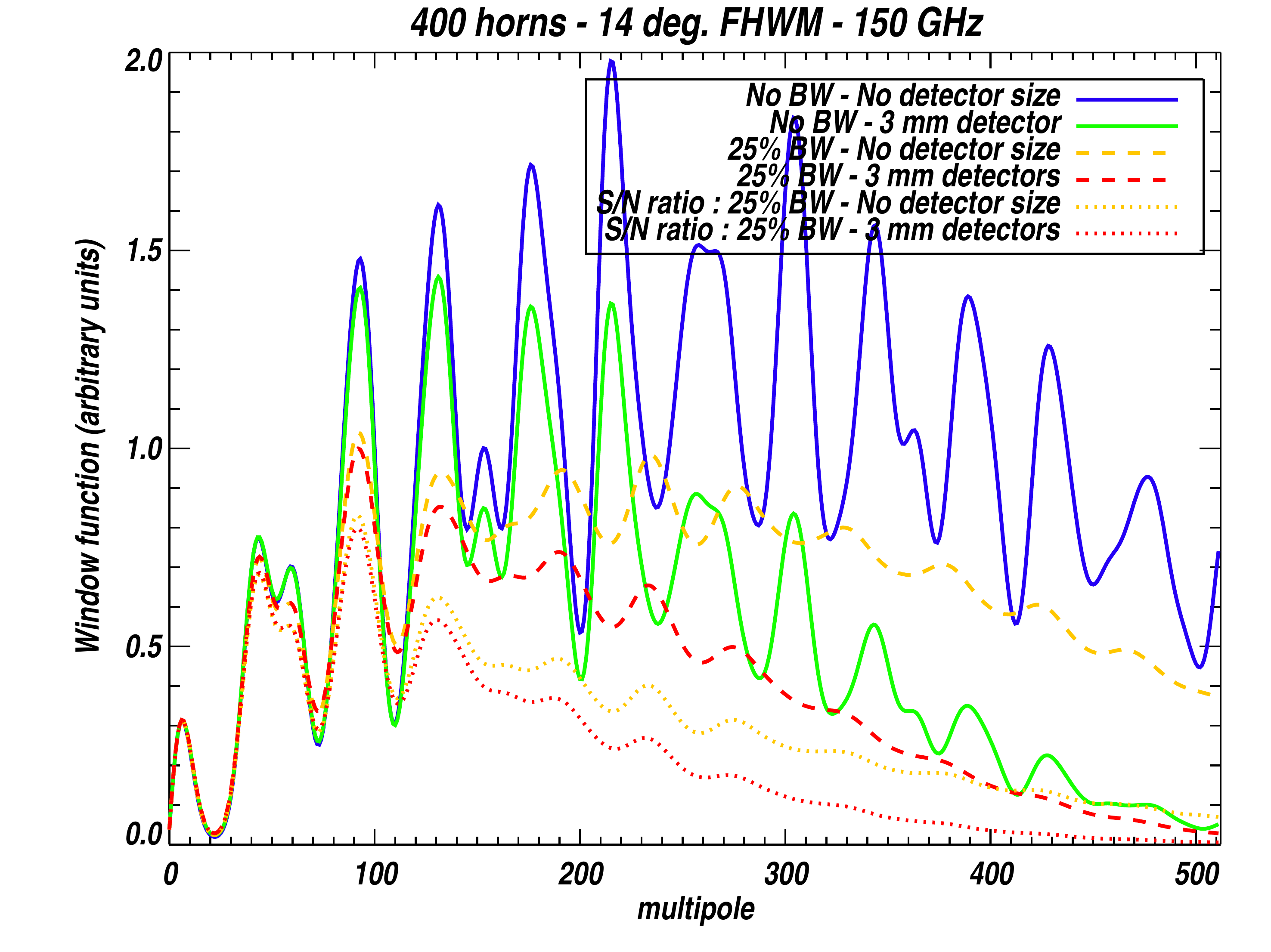}}}
   \caption{Diagonal ($p=p'$, $\vec{n}_k=\vec{n}_l$, $\theta_i=\theta_j$) window functions for QUBIC (in arbitrary units) in the cases of point-like or 3 mm size bolometers, neglecting and accounting for 25\% bandwidth smearing (dashed lines). The two bottom curves (dotted lines) show the evolution of the signal-to-noise ratio expected on the power spectrum accounting also for the noise penalty coming from bandwidth smearing (see Eq.~\ref{kappa1}). Note that above $\ell\sim 200$, the statistical noise becomes dominant although it does not appear on this plot.}
              \label{wfuncts}%
    \end{figure}

It is useful to have a look at the window functions as they give a good feeling of the sensitivity of the instrument to various multipoles. The diagonal parts of the window functions ($p=p'$, $\vec{n}_k=\vec{n}_l$, $\theta_i=\theta_j$) are shown in Fig.~\ref{wfuncts} in the case of either point-like or 3 mm detectors, with or without bandwidth smearing included. 
The peaks that are clearly seen in this figure show that we have access to some discrete sample of all the multipoles with a resolution in multipole space given by the inverse of the primary beam $\Delta\ell \sim 1/\sigma_\mathrm{primary}$. One can actually exactly recover the monochromatic, point-like detector window function just from the number of equivalent baselines at each multipole and this multipole space resolution.
One clearly observes the suppression of power at high multipoles due to integrating the fringes smaller than the size of the detectors as well as the reduction of power and smoothing in multipole space due to bandwidth smearing. The latter is in perfect agreement with analytical calculations performed in visibility space in~(\cite{bandwidth}). One should notice that the bandwidth smearing also causes an additional signal-to-noise reduction (enhancement of the noise due to the reduction of the effective primary beam, see Eq.~\ref{kappa1}) that is not visible in the window function which just shows the amount of signal that is filtered by the instrument. This is detailed in~(\cite{bandwidth}).  For this reason we show in Fig.~\ref{wfuncts} the two bottom curves (dotted lines) which display the actual evolution of the signal-to-noise ratio as a function of $\ell$ due to the full bandwidth effect.

\subsection{Scanning the sky}
Using QUBIC just as an imager is therefore completely equivalent to using it as an interferometer. It actually brings in many advantages through the possibility of scanning the sky, like usual imagers and unlike usual interferometers. 
Scanning the sky actually offers an immense advantage with respect to observing single fields thanks to the extra-modulation of the sky signal it allows. The half-wave plate polarization modulation shown in Eq.~\ref{signal} does not allow recovery of the total power signal, unless one scans the sky, as total power is not modulated by the half-wave plate.  Scanning the sky is therefore the only way of recovering $I$ in addition to $Q$ and $U$. Furthermore (and even more important), spurious signals coming from reflections of unpolarized background light signals on the various components in the optical path, polarized by the polarizing grid, modulated and re-reflected by the half-wave plate, will induce significant disturbances precisely at four times the modulation frequency of the half-wave plate, just where the $Q$ and $U$ sky signals are modulated~(\cite{salatino1}, \cite{salatino2}). This major issue with half-wave plates has been addressed in detail in~(\cite{johnson}) and the removal of this waveplate-synchronous disturbance proved to be only possible through scanning the sky: this scanning slightly shifted the signal modulation frequency sideband from the central $4\omega$, where the disturbance was dominant. In our case, scanning is therefore necessary to extract the synthetic images modulated by the half-wave plate.

\subsection{Scanning and Map-making}
Exactly in the same way as for a usual imager, we can scan the sky with our synthesized beam with any scanning strategy.  Each individual measurement results in the convolution of the sky through our synthesized beam. The Time-Ordered Data of each individual bolometer can then be reprojected on the sky using standard map-making techniques. The usual pseudo-power spectrum estimation can then be applied to the resulting maps. Such simulations are currently ongoing within the QUBIC collaboration.

\begin{figure}[!t]
   \centering\resizebox{\hsize}{!}{\centering{
   \includegraphics{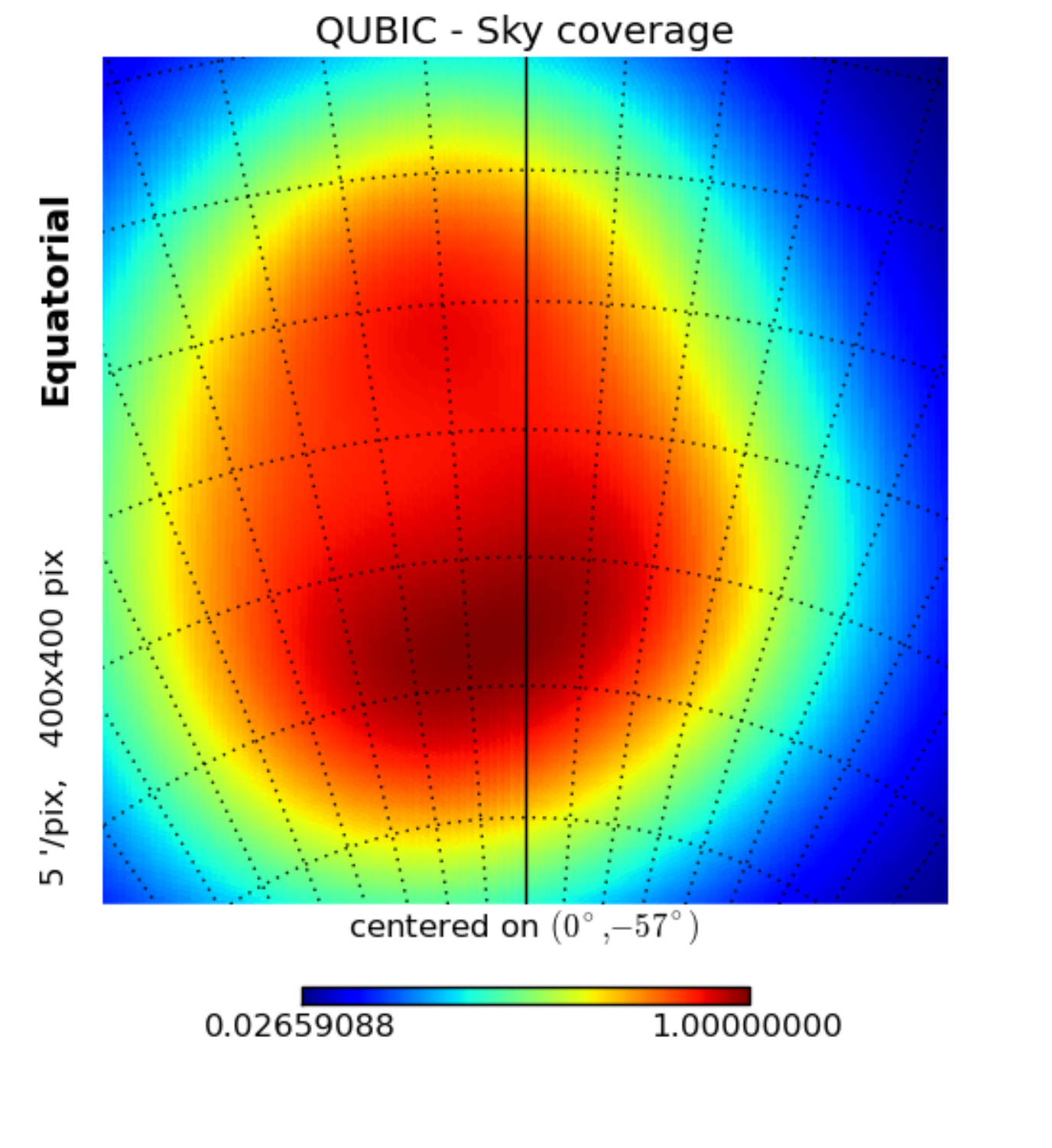}}}
   \caption{Field of view, or sky coverage, obtained with QUBIC by scanning $\sim 10$ degrees in azimuth with slowly varying elevation (from 45 to 65 degrees). The fraction of sky covered is $\sim 2\%$ of the whole sky with a rather flat profile achieving $\eta=1.6$.}
              \label{fov}%
    \end{figure}

We plan to use a scanning strategy based on azimuth scans with $\Delta \mathrm{Az}\sim 10$ degrees slowly varying the elevation (ranging from 45 to 65 degrees) after a number of scans in order to have the center of the field of view scanning an approximate circle on the sky. 
The scanning parameters must be adapted to the instrument design. The
primary beam of QUBIC is around $14$ degrees; the azimuth scans should therefore
be less extended than for traditional imagers, to avoid galactic
contamination. We have chosen a typical azimuth scan of $10$ degrees,
close to the distance between main and secondary peaks of the
synthetic beam. The scanning speed must be fast enough to modulate the $I$
sky signal above the $1/f$-noise: assuming the detectors are stable over
a period of $10\,$s, it requires the scanning speed to be of the order
of $2$ degrees per second. Finally, the half-wave plate should rotate fast
enough to allow measurement of $Q$ and $U$ before the main peak of the synthesized beam moves more than a fraction of a beamwidth.  Given the scanning speed and the
main peak size (around a degree), a rotation frequency of a few Hertz
is enough. These parameters are actually not very far from the ones
chosen for previous imaging experiments (\cite{takahashi},~\cite{johnson}), and so should
be technically achievable.
As a result the final field of view (shown in Fig.~\ref{fov}) defined by the sum of the primary beams along this circle will be significantly flattened and would achieve a value of $\eta =1.6$ (see Fig.~\ref{sensitivity} for an explanation of $\eta$).  

\section{Sensitivity comparison with an imager}
It is of course critical to estimate the sensitivity of an instrument like QUBIC and to compare it to that of an imager with the same number of horns and similar angular resolution. This study has been performed in detail, neglecting the effect of bandwidth, in~(\cite{sensitivity}). A full account of bandwidth smearing and its effect on CMB measurements can be found in~(\cite{bandwidth}) so that one can obtain a general formula for the sensitivity of a bolometric interferometer on the CMB polarization (E or B) power spectrum:                                                                                                                                                                                                                                                                                                                                                                                                                                                                                                                                                                                                                                         
\begin{equation}
\Delta C_\ell^\mathrm{BI} = \sqrt{\frac{2\kappa_1(\ell)}{(2\ell+1) f_\mathrm{sky}\Delta \ell}}
\left(  C_\ell+\frac{2\eta N_h \mathrm{NET}^2 \Omega}{N_\mathrm{eq}^2(\ell) t}\kappa_1(\ell) w^{-1}_\mathrm{pix}(\ell) \epsilon^{-1}_\mathrm{BI}   \right)      \label{sensbi}
\end{equation}
where $t$ is the integration time, $\Omega=2\pi\sigma_\mathrm{beam}^2$ is the solid angle subtended by the primary beam (therefore the field of view of the instrument), $f_\mathrm{sky}=\Omega/2\pi$ is the fraction of the sky covered by the instrument. $N_h$ is the number of horns installed in a compact square array, $N_\mathrm{eq}(\ell)$ is the number of equivalent baselines at multipole $\ell$, $\mathrm{NET}$ is the {\em noise equivalent temperature} of the detectors that scales $\propto 1/\sqrt{\Delta\nu}$. $\eta$ in the above equation stands for the apodization factor of the field of view $\eta=\frac{\int B_\mathrm{prim}(\vec{n})\mathrm{d}\vec{n}}{\int B^2_\mathrm{prim}(\vec{n})\mathrm{d}\vec{n}}$. For a Gaussian primary beam it is equal to 2 but would be one if the field of view had a rectangular shape. By scanning the sky with the bolometric interferometer, we will achieve a value for $\eta$ that will be between 1 (optimistic) and
2 (pessimistic). $\epsilon_\mathrm{BI}$ is the overall optical efficiency of the bolometric interferometer. $w_{pix}(\ell)$ represents the loss in signal to noise due to integration over the detector area (this is the ratio of the blue and green curves in Fig.~\ref{wfuncts}).  $\kappa_1(\ell)$ is a quantity that defines the effect of bandwidth smearing in interferometry (see~\cite{bandwidth} for details), for a Gaussian primary beam and a Gaussian bandwidth with $\Delta\nu = \sigma_\nu \sqrt{2\pi}$, it can be analytically calculated:
\begin{equation}
\kappa_1(\ell) = \sqrt{1+\frac{(\Delta\nu/\nu)^2}{\sigma_\ell^2}\ell^2} \label{kappa1}
\end{equation}
where $\sigma_\ell$ is the resolution of the interferometer in multipole space, $\sigma_\ell=1/\sigma_\mathrm{primary}$. Equation~\ref{sensbi} is the equivalent for bolometric interferometry of the well known Knox formula for imagers~(\cite{knox}):
\begin{equation}
\Delta C_\ell^\mathrm{Im} = \sqrt{\frac{2}{(2\ell+1)f_\mathrm{sky}\Delta_\ell}}\left( C_\ell +\frac{2\eta \mathrm{NET}^2 \Omega}{N_h B_\ell^2 t}\epsilon^{-1}_\mathrm{Im}\right)
\end{equation}
where we assume a uniform sky coverage and the beam window function is given by $B_\ell=\exp\left[-\frac{1}{2}\ell(\ell+1)\sigma_\mathrm{beam}^2\right]$. The parameter $\eta$ corresponds to the sky coverage uniformity ($\eta=1$ for a uniform sky coverage).

\begin{figure}[!t]
   \centering\resizebox{\hsize}{!}{\centering{
   \includegraphics{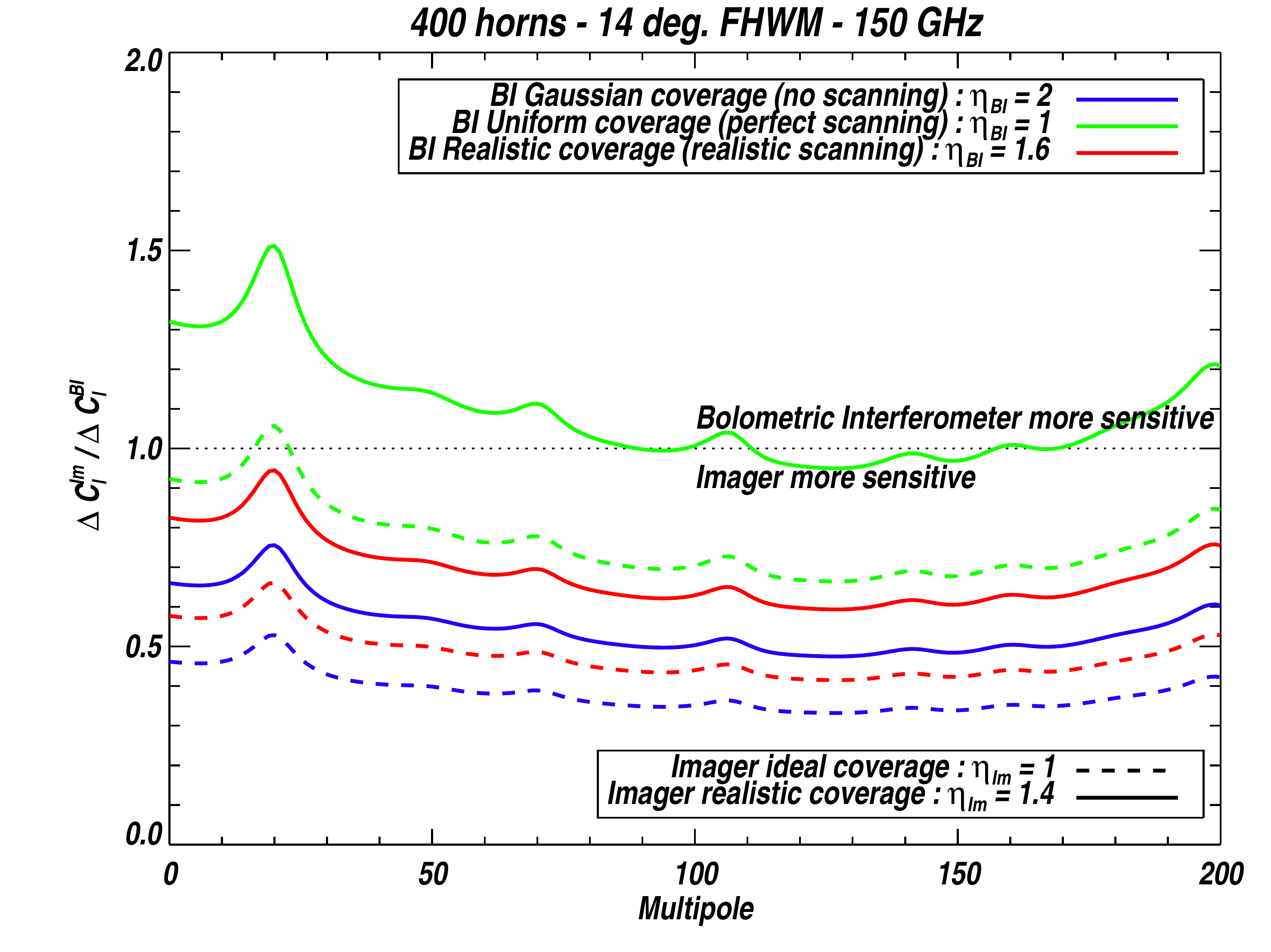}}}
   \caption{Comparison of the sensitivity to the CMB power spectrum of an imager with 1 degree (FWHM) resolution with a bolometric interferometer with a primary beam of 14 degrees (FWHM).  Each system has 400 horns, 25\% bandwidth, and a center frequency of 150 GHz. The interferometer has detectors of size 3 mm.  The dashed curves correspond to an imager with a perfectly uniform sky coverage ($\eta_\mathrm{Im}=1$) while the solid ones are for a more realistic sky apodization including actual scanning strategy and filtering with $\eta_\mathrm{Im}=1.4$ (taken from QUAD -- \cite{brown}). The colors corresponds to various assumptions regarding the bolometric interferometer scanning strategy: single field (no scanning) with $\eta_\mathrm{BI}=2$ is in blue and an unrealistic uniform sky coverage ($\eta_\mathrm{BI}=1$) is in green. The present scanning strategy for QUBIC corresponds to $\eta_\mathrm{BI}=1.6$ is shown in red.}
              \label{sensitivity}%
    \end{figure}

One can therefore calculate the ratio of the noise parts of these equations to have a feeling for the relative sensitivities between an imager and a bolometric interferometer. We will assume that they have the same number of horns and cover the same sky fraction (in one single field for the bolometric interferometer, and by co-adding maps for the imager). The ratio is:
\begin{equation}
\left. \frac{\Delta C_\ell^\mathrm{Im}}{\Delta C_\ell^\mathrm{BI}}\right|_\mathrm{noise} = \frac{\eta_\mathrm{Im}}{\eta_\mathrm{BI}}\times w_\mathrm{pix}(\ell)\times \kappa_1^{-3/2}(\ell) \times \frac{\frac{N_\mathrm{eq}^2(\ell)}{N_h^2}}{B_\ell^2}\times \frac{\mathrm{NET}_\mathrm{Im}^2}{\mathrm{NET}_\mathrm{BI}^2}\times \frac{\epsilon_\mathrm{BI}}{\epsilon_\mathrm{Im}}\label{ratio}
\end{equation}
We show the result of the above ratio in Fig.~\ref{sensitivity} assuming the same NET and optical efficiencies for both instruments. We show different cases for the sky apodizations corresponding to ideal or more realistic scanning strategies for both the imagers and QUBIC. The curve that is the most likely to be realistic is the solid-red one corresponding to a realistic sky apodization ($\eta_\mathrm{Im}=1.4$ for the imager, taken from QUAD -- \cite{brown})\footnote{The value of $\eta_\mathrm{Im}=1.4$ quoted for QUAD corresponds to the result at the end of the data analysis process and therefore includes all filtering. It is  hard to compare fairly with the number we quote for QUBIC that is an anticipation.} and to the scanning strategy presented in Fig.~\ref{fov} for QUBIC with $\eta_\mathrm{BI}=1.6$.
We observe that the bolometric interferometer seems slightly less sensitive than the imager, but the difference is small considering the uncertainties in all the factors used for the comparison. So one can conclude that there is no clear advantage from the statistical sensitivity point of view to any of the two.

\section{Systematic effects\label{syste}}
The expected weakness of the B-mode signal compared to temperature and E-mode anisotropies implies that instruments dedicated to this search not only need to be highly sensitive from a statistical point of view, but they also need to have an exquisite handle on systematic effects that usually cause total power and polarization leakage. This reason is the driver for our efforts to develop a bolometric interferometer as interferometers are known to allow for a very accurate accounting of systematic effects. 

\subsection{Different systematics in interferometry and imaging}
A first point regarding systematic effects is that we anticipate them to be significantly different in our interferometer as compared to all other experiments, most of which are imagers with bolometric detectors. It has been recognized for a long time that systematics effects behave in a significantly different manner with an interferometer than with an imager~(\cite{bunnsyst}). A non-exhaustive list of differences relevant to our case follows:
\begin{itemize}
\item Pointing mismatch and gain differences between detectors that are differenced in order to measure $Q$ and $U$ will not be an issue in our case thanks to the use of the half-wave plate that allows a measurement of all Stokes parameters with one total power detector\footnote{Note that most upcoming imagers dedicated to B-mode search will also use a half-wave plate to modulate the polarization.}.
\item Cross-polarization coming from the telescope in an imager occurs before the half-wave plate and is therefore modulated as the sky signal. In our case, the half-wave plate is located right after the horns and there is no telescope before. We expect the cross-polarization from the window, filters and horns to be very small. The cross-polarization induced by the telescope or the polarizing grid will not be modulated as the sky signal and will therefore not affect the signal in a dangerous manner.
\item The interferometer's scanning strategy can be slow as it does not need to scan a large area of the sky (the primary beam is already large) so long bolometer time constants (that are often far from pure exponentials) will have a lesser impact than for an imager.
\item Ground pickup from sidelobes induced by the telescope have been a strong limitation for imagers from the ground. The interferometer primary beams are determined by horns that have small intrinsic sidelobes.
\item The interferometer's resolution (the synthesized beam) is completely defined by the primary horn array and differences in the primary beams will have a very small impact on the reconstructed signal. The synthesized beam can be calculated in several ways (see below) which can be compared and checked for consistency. The strong limitation in high-multipoles measurements for imagers due to beam uncertainties will therefore be mitigated in our case thanks to the extreme accuracy of our knowledge of the synthesized beam.
\end{itemize}
Having an instrument affected by systematics in a different way as the others seems to be a very important step towards having a convincing detection of the B-mode signal. 

\subsection{Autocalibration}
The main reason to think that systematics might be more controlable in a bolometric interferometer than in a traditional imager is, however, different.  It is related to a specific {\em autocalibration} technique that has been developed in our collaboration~(\cite{autocal} and~\cite{romainphd}).  The autocalibration allows the determination of most of the systematic effects (as modeled using the Jones matrix formalism for instance) independently for all channels from observations of a polarized source (whose polarization does not need to be known) using the switches that can be seen in the schematic view of the instrument  (Fig.~\ref{sketch}) between the back-to-back horns. 

This autocalibration technique is inspired by traditional interferometry (especially long baseline and optical) where the phase is often lost due to atmospheric turbulence (see~\cite{pearsonreadhead} for a review). Two main techniques have been used to solve this problem in the past.  The first technique is based on {\em closure phase} (see for instance~\cite{readheadwilk}) where the unknown phases are iteratively reconstructed by forming quantities where the unknown phases are nulled (for instance the product of the three visibilities one can form with three receivers). The second, more powerful technique is based on the redundancy of the receiver array~(\cite{yang}, ~\cite{wieringa}, ~\cite{noordam}). It is rarely used because most interferometers have low redundancy as they seek high angular resolution and good image reconstruction (dense {\em uv-plane} coverage) rather than sensitivity. The ``omniscope" proposed in~(\cite{omniscopea} and~\cite{omniscopeb}) however relies on redundancy in the same way we do (it is actually a numerical equivalent of our bolometric interferometer, but focused on 21cm observations) and uses them as a source of autocalibration~(\cite{omniscopec}) although they are not sensitive to polarization.

The autocalibration technique uses the fact that in the absence of systematic effects, equivalent baselines of the interferometer should measure exactly the same quantity. Using the switches located between the back-to-back horns one can modulate on/off a single pair of horns while leaving all the others open in order to access the visibility measured by this pair of horns alone. By repeating this with a subset of all available baselines (equivalent and different), one can construct a system of equations whose unknowns are the systematic effects parameters for each channel (as modeled using Jones matrices for instance) meaning two polarizations for each primary horn on each of the bolometers of the focal plane. One can show that for a large enough array of primary horns (at least $\sim 20$ horns) the system is over-constrained and can be solved. No information is required on the actual polarization of the observed source except that it needs to be polarized. The autocalibration requires the knowledge of the individual primary beams of each horn that can be obtained through scanning an external unpolarized source. Once all the couplings and gains are known for each channel and for each bolometer in the focal plane, the synthesized beam can be calculated very accurately.  It can be compared to a direct measurement obtained by opening all the switches and scanning a source as well as to direct calculations using the positions of the horn and the optical simulation of the beam combiner. All these comparisons will allow for a number of consistency checks and measurements of systematic effects that will allow handling these effects in a very precise way. 

\begin{figure}[!t]
   \centering\resizebox{\hsize}{!}{\centering{
   \includegraphics{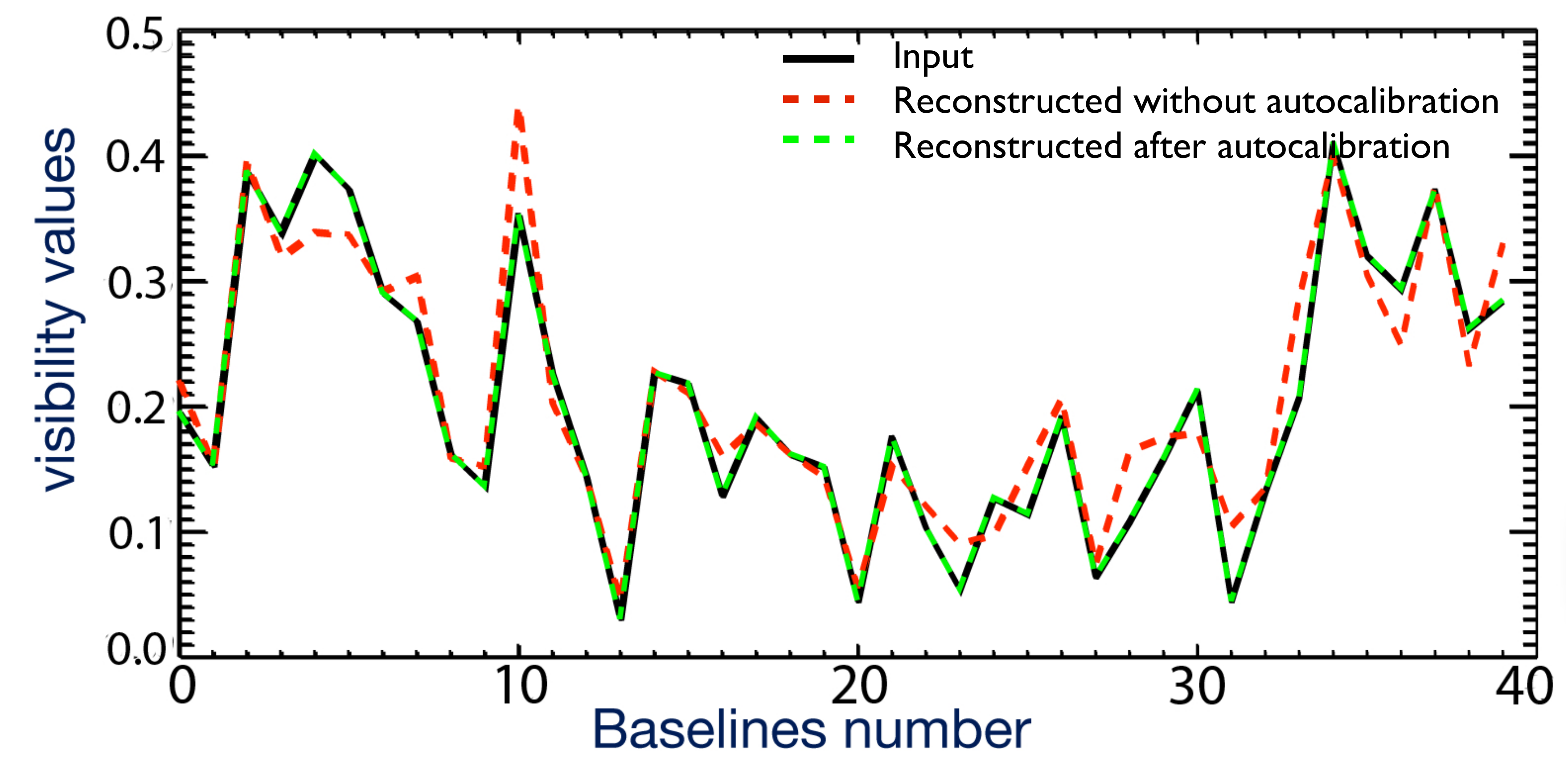}}}
   \caption{Simulation of the autocalibration technique: we show in black the set of visibilities used as an input (the x-axis is just the label of the different visibilities) generated with a 144 horns bolometric interferometer. The dashed red line shows the visibilities reconstructed without accounting for systematics. The corruption is obvious and is completely solved after applying the autocalibration technique (green dashed) that allows to access the systematic coefficients (gains and couplings for each channel).}
              \label{autocalfigure}%
    \end{figure}

We have simulated this autocalibration technique in order to check its behavior (in the absence of noise) and we obtain excellent results as shown in Fig.~\ref{autocalfigure}. We start from a set of visibilities observed with a 144 horns bolometric interferometer and apply randomly drawn systematic effects on each channel (complex gains and coupling using a Jones matrix formalism), the reconstructed visibilities are clearly corrupted (red dashed line) without using the autocalibration technique, but are exactly reconstructed (there is no noise in this simulation) when one includes in the visibility reconstruction the systematics coefficients determined by the autocalibration technique (green dashed line, superimposed to the black one). This shows that the autocalibration techniques actually works fine and allows recovery of the systematics in an efficient manner. An over-simplistic back of the envelope calculation shows that with $\mathrm{NET}\sim300~\mu\mathrm{K.sec^{1/2}}$ detectors and a $\mathrm{T}\sim100$~K polarized source, one can reach an accuracy of order $\mathrm{T/NET}\sim 3\times 10^{-6}$ on each of the Jones matrix coefficient if one spends one second on each baseline. For the whole array this would imply an autocalibration procedure that would last $\sim$ 9 hours. It could be done once in a while and on a more regular basis one could perform the autocalibration only with baselines within subarrays of a smaller number of horns allowing to perform the autocalibration in about one hour.

There is no equivalent to this autocalibration technique with an imager that will just rely on scanning sources (polarized and unpolarized) with the instrument. We can gather more informations than that with our bolometric interferometer and therefore hope that the bolometric interferometer will allow to handle the systematic effects in a more accurate way than an imager, making the best of its interferometric nature.

\section{Conclusions}
We have presented in this article the QUBIC instrument, a bolometric interferometer combining the signals from an array of wide ($\sim 14$ degrees) entry horns using a cold telescope as an optical combiner  to form interference fringes of $I$, $Q$ and $U$ Stokes parameters (modulated with a half-wave plate) on a bolometer array. We have shown that such an instrument can be achieved with technology that is mostly already available and that a small amount of R\&D is required for some of the components, but that are common with future imagers. The image we observe on the focal plane is actually the so-called ``synthetic image" making our bolometric interferometer a synthetic imager in contrast to usual interferometers whose observables are the visibilities (Fourier Transform of the synthetic image). With this synthetic imager the usual imaging techniques can be applied easily such as scanning the sky (in order to improve the sky coverage uniformity and modulate $I$, $Q$ and $U$), map-making and classical power-spectrum estimation. This simplifies the analysis with respect to the visibility-space one that is more difficult to achieve outside the flat-sky approximation. The statistical sensitivity of QUBIC has been shown to be comparable to that of an imager with the same number of horns covering the same sky fraction while we expect the systematic effects to be different from those of an imager. We also propose an autocalibration technique, specific to bolometric interferometry, inspired by the redundancy based traditional interferometry techniques, that will allow to determine accurately the systematic effects (as models as Jones matrices) for each of the channels of the instrument. We anticipate that this will allow QUBIC to achieve a good control of systematics and therefore approach its statistical sensitivity. The full QUBIC instrument will comprise three frequencies and targets to constrain at the 90\% confidence level a tensor-to-scalar ratio of 0.01 with one year of data taking from the Concordia Station at Dôme C, Antarctica.

\begin{acknowledgements}
The authors wish to thank Matthieu Tristram for his valuable help for simulations. This work was supported by Agence Nationale de la Recherche (ANR), Centre National d'Études Spatiales (CNES), Centre National de la Recherche Scientifique (CNRS), Science and Technology Facilities Council (STFC), the National Science Foundation (NSF), theNational Aeronautics and Space Administration (NASA), Programma Nazionale Ricerche in Antartide (PNRA) and Science Foundation Ireland (SFI).
\end{acknowledgements}


\begin{thebibliography}{}
\bibitem[Kovac et al., 2002]{kovac}
J. Kovac et al., Nature, 420 772 (2002).
\bibitem[Readhead et al., 2002]{readhead}
A.C.S. Readhead et al., Science, 306, 836 (2002).
\bibitem[Reichardt et al., 2009]{reichardt}
C.L. Reichardt, ApJ 694, 1200–1219 (2009), arXiv:0801.1491
\bibitem[Takahashi et al., 2008]{takahashi}
Y. Takahashi et al., Proc. SPIE, 7020, 70201 (2008).
\bibitem[The QUAD collaboration, 2009]{quad}
The QUAD collaboration, ApJ 705, 978–999, (2009)
\bibitem[Murphy et al., 2010]{murphy2010} J.A. Murphy et al, Journ. of Inst., Volume 5, Issue 04, pp. T04001 (2010).
\bibitem[O'Sullivan et al., 2008]{osullivan} C. O’Sullivan et al., Infrared Physics \& Technology, 51, 277–286, (2008)
\bibitem[Maffei et al., 2000]{maffei} B. Maffei et al, Int. Journ. of IT and mm waves, 21, (12) 2023-2033, (2000).
\bibitem[Ade  et al., 2009]{ade} P.A. Ade et al.,  20th Int. Symp. on Space THz Tech., Charlottesville, (2009). 
\bibitem[Maffei et al., 2004]{maffei2004} B. Maffei et al, Proceedings of the SPIE, Volume 5498, pp. 812-817 (2004)
\bibitem[Grimes et al., 2009]{grimes2009} Grimes, P.K. et al, Proc. of the 20th Int. Symp. on Space THz Tech., held April 20-22, 2009, in Charlottesville, 2009, p.97.
\bibitem[Tran et al., 2008]{tran} H. Tran et al., Applied optics, 47, 2, p.103, (2008).
\bibitem[Salatino et al., 2010a]{salatino1} M. Salatino, P. de Bernardis, S. Masi, submitted to A\&A, {\tt arXiv:1006.5392v2}
\bibitem[Salatino \& de Bernardis, 2010b]{salatino2} M. Salatino and P. de Bernardis, Proc. Moriond 2010 {\tt 	arXiv:1006.3225v1}
\bibitem[Pisano et al., 2006]{hwp1} G. Pisano et al, Applied Optics, Vol. 45, Issue 27, pp. 6982-6989 (2006)
\bibitem[Pisano et al., 2008]{hwp2} G. Pisano et al, Applied Optics, Vol. 47, Issue 33, pp 6251-6256 (2008)
\bibitem[Pajot et al., 2008]{TES1} F. Pajot et al, Journal of Low Temperature Physics, vol. 151 pp. 513 (2008)
\bibitem[Piat et al., 2008]{TES2} M. Piat et al, SPIE mm \& Submm Det. and Inst. for Astro. pp. 13 (2008)
\bibitem[Voisin et al., 2008]{TES3} F. Voisin et al, Journal of Low Temperature Physics, vol. 151 pp. 1028 (2008)
\bibitem[Pr\^ele et al., 2009]{TES4} D. Pr\^ele et al, IEEE Transactions on Applied Superconductivity (2009)
\bibitem[Charlassier et al., 2009]{muxpaper} R.~Charlassier, et al., Astron. Astrophys. 497-3, 963:971 (2009)
\bibitem[Hyland et al., 2008]{hyland} P. Hyland, B. Follin et E.F. Bunn, MNRAS, 393, Issue 2 pp531-537 (2008).
\bibitem[Gorski et al, 2005]{healpix} K.M. Gorski et al., Astrophys.J. 622 (2005) 759-771.
\bibitem[Bunn et White, 2006]{mosaicking} E.F. Bunn and M. White, ApJ 655, pp 21-29, (2007).
\bibitem[Johnson et al., 2007]{johnson} B.R. Johnson et al., APJ  665. issue 1, pp 42-54  (2007).
\bibitem[Bunn et al., 2003]{bunn_zald} E.F. Bunn et al., Phys.Rev.D67:023501, (2003)
\bibitem[Charlassier et al., 2010]{bandwidth} R. Charlassier et al, A\&A 514, A37,  (2010)
\bibitem[Hamilton et al, 2008]{sensitivity} J.-Ch. Hamilton et al., A\&A 491, 923-927 (2008)
\bibitem[Knox, 1997]{knox} L.~Knox, ApJ, v480, p72, {\tt arXiv:astro-ph/9606066} (1997).
\bibitem[Brown, priv. comm.]{brown} M. Brown, private communication.
\bibitem[Charlassier et al., in prep.]{autocal} R. Charlassier et al, in preparation.
\bibitem[Charlassier , 2010]{romainphd} R. Charlassier, PhD, University Paris-Diderot, 2010.
\bibitem[Bunn, 2006]{bunnsyst} E.F. Bunn, Phys.Rev.D75:083517(2007).
\bibitem[Pearson and Readhead, 1984]{pearsonreadhead} T.J. Pearson and A.C.S. Readhead, Ann. Rev. A\&A, 22:97-130 (1984).
\bibitem[Readhead and Wilkinson, 1978]{readheadwilk} A.C.S Readhead and P.N. Wilkinson, ApJ 223:25-36 (1978) 
\bibitem[Yang, 1988]{yang} Y.P. Yang, A\&A 189:361-364 (1988).
\bibitem[Wieringa, 1991]{wieringa} M. Wieringa, Ast. Soc. of the Pacific Conf. series, vol. 19:192-196 (1991).
\bibitem[Noordam and de Bruyn, 1982]{noordam} J.E. Noordam and A.G. de Bryn, Nature, 299:597 (1982).
\bibitem[Tegmark and Zaldarriaga, 2009a]{omniscopea} M. Tegmark and M. Zaldarriaga, Phys.Rev.D 79(8):03530 (2009).
\bibitem[Tegmark and Zaldarriaga, 2009b]{omniscopeb} M. Tegmark and M. Zaldarriaga, {\tt arXiv:0909.0001v1}.
\bibitem[Liu et al, 2010]{omniscopec} A. Liu et al., {\tt arXiv:1001.5268v3}
\end{thebibliography}
\end{document}